\documentclass[twocolumn, longbibliography, prr, aps, amsmath, amssymb,superscriptaddress]{revtex4-2}
\usepackage{amsmath}
\usepackage{amsfonts}
\usepackage{graphicx}
\usepackage[colorlinks,linkcolor=blue,citecolor=blue,urlcolor=blue]{hyperref}
\usepackage{float}
\usepackage[normalem]{ulem}
\usepackage[switch,columnwise]{lineno}
\usepackage[dvipsnames]{xcolor}

\begin{document}
	\title{Weak Ergodicity Breaking in Optical Sensing}
	
	\author{V. G. Ramesh}
	\author{S. R. K. Rodriguez}
	\affiliation{Center for Nanophotonics, AMOLF, Science Park 104, 1098 XG Amsterdam, Netherlands\\}
	
	\date{\today}
	
	\begin{abstract}
		The time-integrated intensity transmitted by a laser driven resonator obeys L\'evy's arcsine laws [Ramesh \textit{et al.}, Phys. Rev. Lett. \textit{in press} (2024)]. Here we demonstrate the implications of these laws for optical sensing. We consider the standard goal of resonant optical sensors, namely to report a perturbation to their resonance frequency. In this context, we quantify the sensing precision attained using a finite energy budget combined with time or ensemble averaging of the time-integrated intensity. We find that ensemble averaging outperforms time averaging for short measurement times, but the advantage disappears as the measurement time increases. We explain this behavior in terms of weak ergodicity breaking, arising when the time for the time-integrated intensity to explore the entire phase space diverges but the measurement time remains finite. Evidence that the former time diverges is presented in first passage and return time distributions. Our results are relevant to all types of sensors, in optics and beyond, where stochastic time-integrated fields or intensities are measured to detect an event. In particular,  choosing the right averaging strategy can improve sensing precision by orders of magnitude with zero energy cost.
	\end{abstract}
	\maketitle

\begin{figure}
		\centering
		\includegraphics{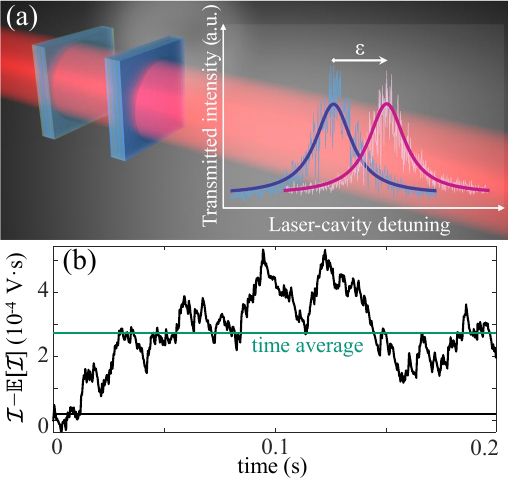}
		\caption{\label{fig_1} (a) A laser-driven optical cavity is used as a sensor. By measuring the cavity transmission, a perturbation $\epsilon$ to the resonance frequency can be detected. (b) Sample trajectory of the time-integrated intensity $\mathcal{I}$, relative to its expectation value $\mathbb{E}[\mathcal{I}]$, measured with our experimental tunable cavity setup as explained in the text.}
  \label{fig1}
	\end{figure}

Optical sensors play a major role in physics research and technology. They are used to measure nanoparticles~\cite{Zhu10, Kelkar15, Zhi17}, single molecules~\cite{Vollmer08, Subramanian18}, chirality~\cite{Feis20}, weak forces~\cite{Guha20}, nanoscale chemical reactions~\cite{Oksenberg21}, distances~\cite{Yang20}, fields~\cite{Lewis20}, and gravitational waves~\cite{Abbott16}, to name a few examples. A popular type of sensor is a laser-driven resonator, or cavity as illustrated in Fig.~\ref{fig1}(a). When the object of interest is close by, the state of the cavity changes. This change is usually detected by integrating the transmitted intensity in time using a photodetector. However, the noise is also integrated. Consequently, the sensing precision that can be attained with limited measurement time and optical power is finite. To maximize this precision, measurements can be averaged in time  or over an ensemble of independent and identically distributed ($iid$) samples. The latter can be achieved by using multiple cavities instead of one, all probing the same environment. In this context, we are interested in the following question: Given an energy budget to estimate the mean time-integrated intensity, will one measurement of duration $\tau_m$ or $m$ $iid$ measurements of duration $\tau_m/m$ yield a better estimate? For an ergodic process, both measurements should give equally precise estimates. However, as we will show, ergodicity is not a given in finite-time sensing.


Ergodicity is defined by the hallmark property of equal time and ensemble averages. Whether and how ergodicity holds has been a central problem in statistical mechanics since its inception and throughout its history~\cite{Lebowitz, Moore15}. Ergodicity can be broken in two ways. The first, known as strong ergodicity breaking, occurs in systems whose phase space is divided such that it cannot be accessed entirely in a single trajectory. The second, known as  weak ergodicity breaking (WEB), occurs in systems whose phase space is unrestricted but the time spent in any region diverges. The distinction between strong and weak ergodicity breaking was introduced in the context of spin glasses~\cite{Bouchaud1992}, where it remains an active research topic~\cite{Bernaschi20,Cugliandolo95,Folema23}. In addition, WEB has drawn interest in various contexts, such as of quantum thermalization~\cite{Polkovnikov11,Russomanno22,Surace20,Turner18,Schecter19,Serbyn21}, biological~\cite{Lomholt07,Jeon11,Manzo15} and neurological processes~\cite{Weron2017}, stochastic resetting~\cite{Stojkoski22}, heterogeneous diffusion~\cite{Singh22,Cherstvy15}, quantum emitters~\cite{Margolin05,Margolin06}, single particle tracking~\cite{Burov11}, and finance~\cite{Stojkoski2022}.

In this manuscript we demonstrate WEB in optical sensing.  We present experimental and numerical results for a laser-driven linear optical cavity, which can perform sensing as illustrated in Fig.~\ref{fig1}(a). The time-integrated intensity transmitted by this system obeys the arcsine laws as shown in our companion Letter~\cite{comp}, and WEB can therefore be expected~\cite{Bel06,   Rebenshtok08, Korabel13}. Here we investigate the consequences of this WEB for finite-time sensing. We analyze the sensing precision when time and ensemble averaging are performed with a fixed energy budget. We identify two distinct ensemble averages to which the time average can be fairly compared. One uses equal power, the other uses equal measurement time, and both use equal amount of energy as the time average. We find that the ensemble average with equal power as the time average is vastly superior for short measurement times. However, as the measurement time increases,  differences between the three averages disappear and ergodicity emerges. To explain our findings, we analyze first passage time and return time distributions of the time-integrated intensity. These distributions have power-law tails, implying arbitrarily long times for the time-integrated intensity to explore its entire phase space and thus for ergodicity to emerge. Finally, we discuss hardware considerations for utilizing our findings, the generality of our results, and their relevance to other types of sensors or time-integrated signals in optics and beyond.

Figure~\ref{fig1}(a) illustrates the system under study. It is a single-mode linear cavity driven by a noisy laser. In a frame rotating at the laser frequency $\omega$, the intra-cavity field $\alpha$ satisfies	
	\begin{equation}
		i\dot\alpha = \left(-\Delta- i\frac{\Gamma}{2}\right)\alpha + i\sqrt{\kappa_L}A + \frac{D}{\sqrt{2}}\left(\zeta_R(t) + i\zeta_I(t)\right).
		\label{eq1}
	\end{equation}
\noindent$\Delta=\omega-\omega_0$ is the frequency detuning between the laser and the cavity resonance $\omega_0$. $\Gamma = \gamma_a +\kappa_L+\kappa_R$ is total dissipation rate, with $\gamma_a$ the absorption rate and $\kappa_{L,R}$ input-output rates through the left or right cavity mirror. $A$ is laser amplitude, which we assume to be real. $\zeta_R(t)$ and $\zeta_I(t)$  account for Gaussian white noise in the laser amplitude and phase. These stochastic terms have mean $\langle\zeta_R(t)\rangle =\langle\zeta_I(t)\rangle=0$ and correlation $\langle\zeta_j(t')\zeta_k(t)\rangle = \delta_{j,k} \delta(t'-t)$. Since $\zeta_{R,I}(t)$ are additive noises, have unit variance, and are multiplied by $D/\sqrt{2}$,  the standard deviation of the laser noise is $D$.
	
Equation~\ref{eq1} captures the physics of a single linear optical mode in diverse systems. Here, and in our companion Letter~\cite{comp}, we use it to describe a plano-concave Fabry-P\'erot microcavity as previously realized in~\cite{Hunger10, Trichet14, Vallance16}. However, the physics is exactly the same for whispering-gallery-mode~\cite{Vollmer08}, ring~\cite{Bogaerts12}, photonic crystal~\cite{Pitruzzello18} or any type of resonator where one mode is sufficiently well isolated spectrally and spatially from all other modes. These types of cavities, or resonators, are widely used for optical sensing. Typically, the cavity transmission or reflection is monitored in search for signatures of the object of interest. The object of interest, which we call `the perturbation', usually perturbs the cavity by shifting the spectral lineshape as illustrated in Fig.~\ref{fig_1}(a). Thus, the frequency magnitude of the perturbation $\epsilon$ can be detected by measuring the spectrum.


Since noise is inevitable and measurements always take finite time,  a common approach to detect the perturbation is by analyzing the time-integrated intensity $\mathcal{I} = \int_{0}^{\tau_m} \kappa_R |\alpha(t)|^2 dt$. $\tau_m$ is the measurement time.  For illustration purposes, in Fig.~\ref{fig_1}(b) we plot one trajectory of $\mathcal{I}$ obtained from our tunable cavity setup.  This and all experimental results in this manuscript were obtained by periodically modulating the cavity length across a resonance as shown in Fig.~\ref{fig_1}(a), such that $\Delta(t)=\Delta(t+T)$ with $T$ the period. Then, we constructed a trajectory of $\mathcal{I}$ by successively integrating the transmitted intensity up to time $\tau_m$.

We are interested in three distinct averages of $\mathcal{I}$:

	\begin{subequations}
		\begin{align}
			\langle\mathcal{I}(t)\rangle &= \frac{1}{m}\sum_{j=1}^{m}\mathcal{I}_j(t) , \label{eq2a}\\ 	
			\overline{\mathcal{I}(t)} &= \frac{1}{t}\int_0^{t} \mathcal{I}(s) ds , \label{eq2b}\\
            \mathbb{E}[\mathcal{I}] &= \int \mathcal{I}\mathcal{P}(\mathcal{I}) d\mathcal{I}. \label{eq2c}
	  \end{align}
	 \end{subequations}

\noindent $\langle\mathcal{I}\rangle$  is the ensemble average,  obtained from $m$ samples. $\overline{\mathcal{I}}$ is the time average, obtained from a single measurement of duration $\tau_m$. $\mathbb{E}[\mathcal{I}]$  is the expectation value or statistical average of $\mathcal{I}$, calculated from the stationary probability density function (PDF) of $\mathcal{I}$, namely $\mathcal{P}(\mathcal{I})$. By writing the time dependence of $\langle\mathcal{I}(t)\rangle$ and $\overline{\mathcal{I}(t)}$ explicitly in equations~\ref{eq2a} and~\ref{eq2b} we highlight that averages are done over finite times.

Figure~\ref{fig_1}(b) shows that $\mathcal{I}$ remains above $\mathbb{E}[\mathcal{I}]$ throughout most of its trajectory. This results in an extreme deviation of $\overline{\mathcal{I}}$ from  $\mathbb{E}[\mathcal{I}]$, quantified by the distance between the green and black horizontal lines. Such extreme deviations are not particular to the trajectory we have chosen to display. On the contrary, extreme deviations from the expectation value are likely to occur in stochastic processes governed by the arcsine laws~\cite{Levy40, Barato18, Majumdar20}, such as $\mathcal{I}$~\cite{comp}.

While not apparent in Fig.~\ref{fig1}(b), $\mathbb{E}[\mathcal{I}]$ is not constant in time. Since $\Delta$ is modulated, $\mathbb{E}[\mathcal{I}]$ is also modulated with the same period $T$. The stochastic process under consideration is therefore temporally inhomogeneous. However, this is not crucial to our results. The  arcsine laws also emerge in the absence of a protocol, meaning for temporally homogeneous processes.  Time integration is the key to the arcsine laws. It transforms a Gaussian process (laser noise) into a Wiener process (stochastic time-integrated intensity).  Our companion Letter~\cite{comp} discusses further cases leading to the arcsine laws, as well as signatures of scaling and universality in their emergence.

	\begin{figure*}
		\centering
		\includegraphics{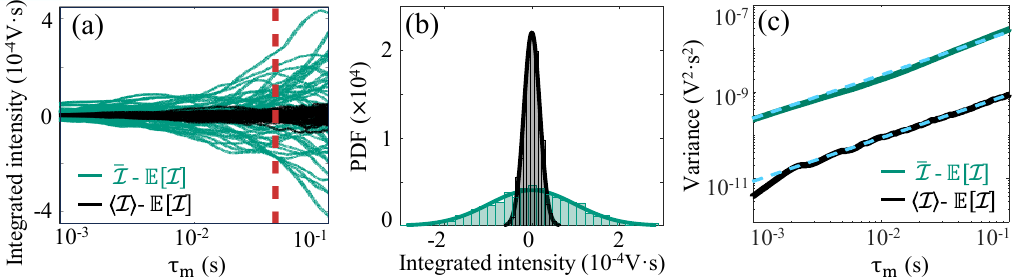}
		\caption{Green and black curves in all panels correspond to the time-integrated intensity when performing time and ensemble averaging,  $\overline{\mathcal{I}}$  and  $\langle\mathcal{I}\rangle$  respectively. The expectation value  $\mathbb{E}[\mathcal{I}]$ is subtracted from both averages.  (a) Experimental trajectories as a function of the measurement time $\tau_m$. (b) Probability density functions (PDFs) at $\tau_m=30$ ms, indicated by the red dashed line in (a). Rectangles are experimental data, solid curves are Gaussian fits. (c) Experimental variance versus  $\tau_m$, with linear fits as blue dashed lines.}
			\label{fig_2}
	\end{figure*}


Let us analyze the statistical properties of  $\langle\mathcal{I}(t=\tau_m/m)\rangle$ and $\overline{\mathcal{I}(t=\tau_m)}$, displayed by our experimental data in Fig.~\ref{fig_2}. The expectation value $\mathbb{E}[\mathcal{I}]$ has been subtracted from both averages for ease of visualization. Figure~\ref{fig_2}(a) shows several trajectories of $\langle\mathcal{I}(t=\tau_m/m)\rangle$ and $\overline{\mathcal{I}(t=\tau_m)}$ versus the measurement time $\tau_m$. Figure~\ref{fig_2}(b) shows the corresponding PDFs at $\tau_m=30$ ms. Already in Figs.~\ref{fig_2}(a,b)  we can see substantial differences in the width of PDFs. We quantify these differences by plotting the variances in Fig.~\ref{fig_2}(c). Both variances grow linearly with the measurement time, as expected for a diffusive L\'evy process such as  $\mathcal{I}$~\cite{comp}. The PDFs and variances in Figs.~\ref{fig_2}(b,c) were obtained
from a much larger data set than the one presented in Fig.~\ref{fig_2}(a) for illustration purposes only. Furthermore, $\langle\mathcal{I}(t=\tau_m/m)\rangle$ and $\overline{\mathcal{I}(t=\tau_m)}$ were obtained from the same data set, involving a single laser-driven cavity. To mimic ensemble averaging with our single cavity, we leveraged the fact that the dynamics are Markovian, meaning memoryless. This allows us to partition a trajectory of duration $\tau_m$ into $m$ trajectories of duration $\tau_m/m$, and to regard the resulting parts as $iid$ samples.


Since $\langle\mathcal{I}\rangle$ has a smaller variance than $\overline{\mathcal{I}}$, we expect ensemble averaging to outperform time averaging in sensing.  To test this expectation, we performed numerical simulations using Equation~\ref{eq1}. We first calculated the dynamics of $\alpha$ for an unperturbed cavity driven on resonance, such that $\Delta=0$.  Then, we set $\Delta=0.1 \Gamma$ (a perturbation of strength $\epsilon= 0.1 \Gamma$ on the cavity) and simulated the dynamics again. The change in $\Delta$ results in a shift in $\mathbb{E}[\mathcal{I}]$, which we aimed to detect by calculating time and ensemble averages of $\mathcal{I}$.

Recognizing energy as the fundamental resource for sensing, we can identify two distinct ensemble averages to which the time average can be fairly compared. To see this, suppose that we determine $\overline{\mathcal{I}}$ using data from one cavity driven with power $A^2$ during time $\tau_m$. Suppose that
we also have $m$ cavities at our disposal, all giving $iid$ measurements of $\mathcal{I}$. To keep the total energy expenditure $m\tau_mA^2$ constant, we can divide the measurement time or the power by $m$.  Therefore, in one scenario the $m$ cavities are driven with equal power $A^2$ for a reduced time $\tau_m/m$. We  denote the ensemble average obtained in this first scenario $\langle\mathcal{I}\rangle_P$, with the `$P$' standing for equal power. In a second scenario, the $m$ cavities are driven with reduced power $A^2/m$ for a time $\tau_m$. We denote the ensemble average obtained in this second scenario $\langle\mathcal{I}\rangle_t$, with the `$t$' standing for equal time. We show below that, while the total energy expenditure is the same for all three averages,  the  sensing precision that can be attained is not.

 \begin{figure*}
		\centering
		\includegraphics{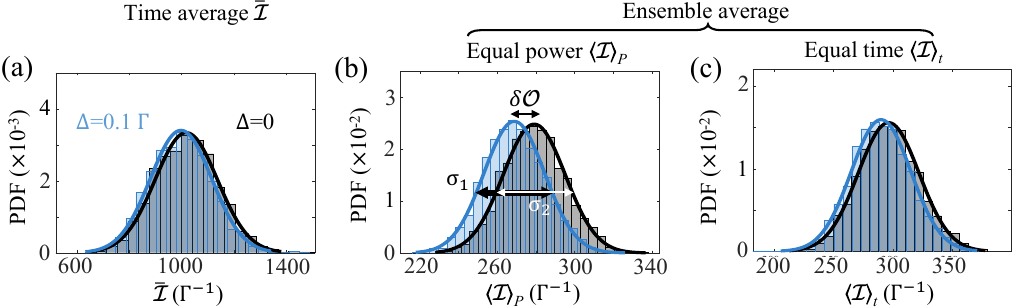}
		\caption{The two probability density functions (PDFs) in every panel are obtained numerically, by calculating the time-integrated intensity $\mathcal{I}$ for two distinct laser-cavity detunings: $\Delta = 0$ in black, and $\Delta = 0.1\Gamma$ in blue. Rectangles are obtained from numerical simulations, and solid curves are Gaussian fits.  In (a) we calculate the time average $\overline{\mathcal{I}}$ of the time-integrated intensity transmitted by a single cavity up to a time $\tau_m=10\Gamma^{-1}$.  In (b) and (c) we calculate ensemble averages of the time-integrated intensity transmitted by 10 cavities. In both (b) and (c), the energy budget (time-power product) is the same as in (a). However, in (b) we consider ensemble averaging with equal laser power as the time average in (a); we call the resultant quantity the equal power ensemble average $\langle\mathcal{I}\rangle_P$. In (c) we consider ensemble averaging with equal measurement time the time average in (a); we call the resultant quantity the equal time ensemble average $\langle\mathcal{I}\rangle_t$. $\sigma_1$ and $\sigma_2$ are the standard deviations of the PDFs. Simulation model parameters: $D=2\Gamma$, $A=10\sqrt{\Gamma}$ in (a) and (b), $A=\sqrt{10\Gamma}$ in (c).
			\label{fig_3} }
	\end{figure*}

We quantify the sensing precision $\chi$ as follows:
    \begin{equation}
        \chi=\frac{\delta\mathcal{O}}{\sqrt{\sigma_1^2+\sigma_2^2}}. \label{eq3}
    \end{equation}
\noindent $\delta\mathcal{O}$ is the shift in the observable $\mathcal{O}$ due to the object of interest. In our case, it is the shift in $\mathbb{E}[\mathcal{I}]$ due to the perturbation $\epsilon$ to the laser-cavity detuning divided by the loss rate $\Delta/\Gamma$.  $\sigma_1$ and $\sigma_2$ are the standard deviations of the PDFs of $\mathcal{O}$ in the presence and absence of the perturbation, respectively.

Figure~\ref{fig_3} illustrates our approach to determine the sensing precision for each of the three aforementioned averages based on numerical simulations. The total energy expenditure $m\tau_mA^2$ is constant. Black and blue bars are PDFs obtained for the unperturbed and perturbed cavity, respectively.  Solid curves of the same color are Gaussian fits to the numerical data. Figure~\ref{fig_3}(a) corresponds to the time average $\overline{\mathcal{I}}$ of the time-integrated intensity transmitted by a single cavity driven with power $A^2=100\Gamma$ during time $\tau_m=10\Gamma^{-1}$. Figures~\ref{fig_3}(b) and  ~\ref{fig_3}(c) correspond to ensemble averages with equal power and equal time, $\langle\mathcal{I}\rangle_P$  and $\langle\mathcal{I}\rangle_t$ respectively, as in the time average. For all three averages, the small $\delta \mathcal{O}$ and large $\sigma_{1,2}$ results in $\chi<1$.  A successful detection strategy can still be constructed in each case with a judiciously defined threshold in $\mathcal{O}$~\cite{Kay}. However, the probabilities of false alarm and missed detection will be significantly smaller for  $\langle\mathcal{I}\rangle_P$  than for the other two averages. In this sense,  the equal power ensemble average is the superior sensing strategy. Arguably, the lower precision of $\langle\mathcal{I}\rangle_t$ than of $\langle\mathcal{I}\rangle_P$ is expected because only for $\langle\mathcal{I}\rangle_t$ we divided the power by $m$ while the noise strength remained constant. However, this argument is insufficient to explain the time dependence of $\chi$ presented next.

	\begin{figure}
		\centering
		\includegraphics{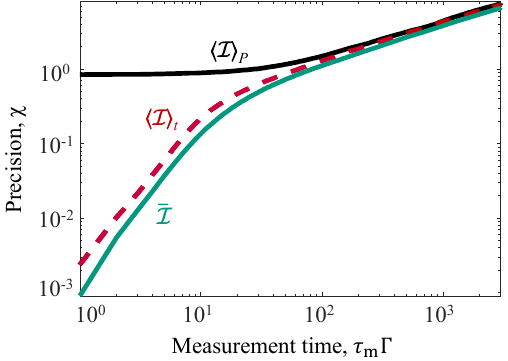}
		\caption{Sensing precision $\chi$ (see Eq.~\ref{eq3}) versus measurement time $\tau_m$, calculated numerically using the model parameters reported in the Fig.~\ref{fig_3} caption. The solid green curve is obtained by time averaging. The solid black and dashed red curves are obtained by ensemble averaging with the equal power and equal time, respectively, as in time averaging. In all three cases, the energy budget (time-power product) is the same.
	\label{fig_4} }
	\end{figure}

Figure~\ref{fig_4} shows $\chi$  versus $\tau_m$ for each of the three averages under consideration. To obtain these results, we used Equation~\ref{eq3} with $\delta \mathcal{O}$  and $\sigma_{1,2}$ evaluated as in the example in Fig.~\ref{fig_3}(b). Again, the total energy expenditure is constant. We first note that all three averages grow with $\tau_m$. As expected, longer measurements give greater precision regardless of the averaging method.  However, important differences between the three averages can be observed for sufficiently small  $\tau_m$. For $\tau_m \sim \Gamma^{-1}$,  the precision of $\langle\mathcal{I}\rangle_P$ exceeds the precision of $\overline{\mathcal{I}}$ by nearly three orders of magnitude. Meanwhile, $\langle\mathcal{I}\rangle_t$ gives roughly a factor of $2$ improvement in sensing precision over $\overline{\mathcal{I}}$. The differences in precision disappear as $\tau_m$ increases, since each average grows at a different rate. Then, in the limit $\tau_m\to\infty$ corresponding to an infinitely long measurement, the precision attained by each average is the same.  Overall, Figure~\ref{fig_4} shows the important consequences of WEB for finite-time optical sensing, and the emergence of ergodicity in sensing at long times.

WEB as observed in Figure~\ref{fig_4} arises from the separation of two timescales. The measurement time $\tau_m$ is finite, but the time required for the underlying stochastic process to fully explore its phase space diverges.  This separation of timescales vanishes for  $\tau_m\to\infty$. Consequently,  for $\tau_m\to\infty$ ergodicity is recovered and  all time and ensemble averages at constant energy give equal precision in sensing. The asymptotic equivalence of time and ensemble averages is a feature specific to weak (rather than strong) ergodicity breaking.  WEB in the time-integrated intensity can be intuitively understood based on the implications of the first arcsine law, as explained in the companion Letter~\cite{comp}. According to the first arcsine law, most  trajectories of $\mathcal{I}$ evolve far above or far below $\mathbb{E}[\mathcal{I}]$. This extreme behavior arises from the properties of a continuous diffusive L\'evy process, such as the time-integrated intensity. Below, we present signatures of such a processes in  first passage time and return times distributions. The  experimental data we present was obtained using the same tunable cavity setup giving the results in Figs.~\ref{fig1} and~\ref{fig_2}.


	\begin{figure}
		\centering
		\includegraphics{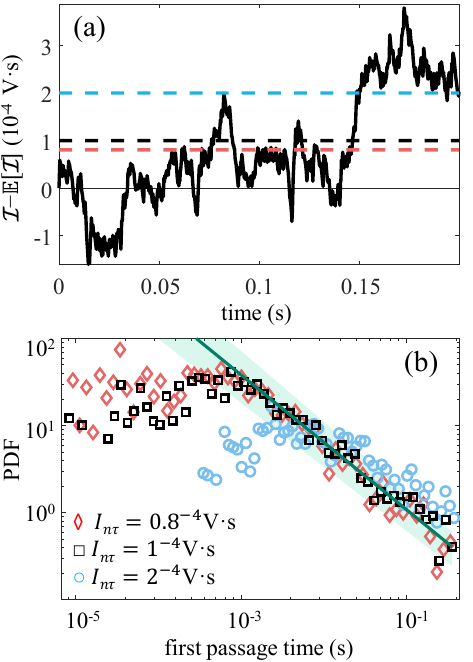}
		\caption{(a) Experimental sample trajectory of the time-integrated intensity $\mathcal{I}$ relative to its expectation value $\mathbb{E}[\mathcal{I}]$.  (b) Red diamonds, black squares, and blue circles are distributions of the times  taken for $\mathcal{I}-\mathbb{E}[\mathcal{I}]$ to cross the dashed lines in (a) of the same color.  The green line is a power law fit with exponent $-1.54$ to the tail of the black curve (cutoff at $\sim10^{-3}$ s). The shaded area is the 95\% confidence interval of the fit.
			\label{fig_5} }
	\end{figure}


In Fig.~\ref{fig_5} we analyze first passage times (FPTs) in trajectories of $\mathcal{I}$. Figure~\ref{fig_5}(a) illustrates how to identify FPTs. Essentially, the FPT is the time $\mathcal{I}$ takes to reach a certain value. Conversely, this may be interpreted as the waiting or occupation time within a given boundary. For concreteness but without loss of generality, we consider three values of $\mathcal{I}$: $0.8^{-4}$ V$\cdot$s, $1^{-4}$ V$\cdot$s, and $2^{-4}$ V$\cdot$s, indicated by the red, black, and blue dashed horizontal lines in  Fig.~\ref{fig_5}(a), respectively. The noisy black curve is an arbitrarily selected trajectory of $\mathcal{I}$. For ease of visualization, we subtracted the expectation value $\mathbb{E}[\mathcal{I}]$ from $\mathcal{I}$. Based on many such trajectories of $\mathcal{I}$, we calculated distributions of FPTs. The results are shown  in Fig.~\ref{fig_5}(b), using the same colour scheme as in Fig.~\ref{fig_5}(a). Notice in Fig.~\ref{fig_5}(b) that the long-time tails of all FPT distributions are linear in the log-log plot, meaning they are characterized by a power law. In each case, we retrieved the corresponding exponents by fitting a power law to the data. For example, for the intermediate value of  $\mathcal{I}$ (black data points) we retrieved an exponent of $-1.54$; the corresponding power law is shown as a dark green line Fig.~\ref{fig_5}(b), and  the light green shaded region is the 95\% confidence interval of the fit. Notice that this power law captures reasonably well (within the uncertainty) the FPT distribution tails for the two other values of $\mathcal{I}$ we selected. We furthermore verified that the FPT distribution tail for any value of $\mathcal{I}$ is well captured by such a power law. The exponent that we find, $-1.54$, is within the uncertainty of the value $-1.5$ expected for a Wiener process, which is a continuous diffusive L\'evy processes~\cite{feller57,bhattacharya90}.

	\begin{figure}
		\centering
		\includegraphics{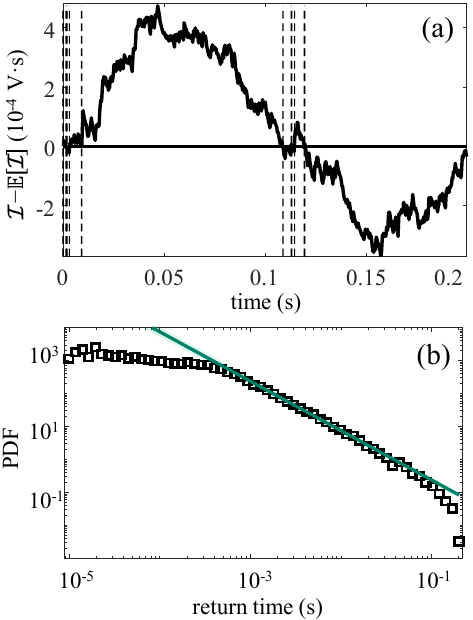}
		\caption{(a) Experimental sample trajectory of  the time-integrated intensity $\mathcal{I}$ relative to its expectation value $\mathbb{E}[\mathcal{I}]$. Dashed lines mark crossings of $\mathbb{E}[\mathcal{I}]$. (b) Probability density function (PDF) of return times, which are the times between two successive crossings of \textbf{$\mathbb{E}[\mathcal{I}]$}. The green line is a power law fit with exponent $-1.5$ (cutoff at $\sim10^{-4}$ s). The light green shaded area is the 95\% confidence interval of the fit.
			\label{fig_6} }
	\end{figure}

In Fig.~\ref{fig_6} we analyze the distribution of return times (RTs) of $\mathcal{I}$. The RT is the time $\mathcal{I}$ takes to return to the expectation value $\mathbb{E}[\mathcal{I}]$. To illustrate the approach, Fig.~\ref{fig_6}(a) shows a different arbitrarily-selected trajectory of  $\mathcal{I}$, from which we have again subtracted $\mathbb{E}[\mathcal{I}]$ for ease of visualisation. The successive returns of this trajectory to the mean are shown by black dashed-lines. Based on many of such trajectories, we calculate the distribution of RTs shown in Fig.~\ref{fig_6}(b). The dark green line is a power law fit of the tail, with cutoff determined by using the Matlab change point detection algorithm. The light-green shaded region (barely wider than the line) is the 95\% confidence interval of the fit. The fitted exponent is $-1.5$, as expected for the tail of distribution of RTs for a Wiener process~\cite{Mingzhou95}.

Probability distributions with power-law tails are characterized by a divergent first moment. This means that the mean FPT and RT are both infinite. From the divergent mean FPT it follows that $\mathcal{I}$ can take arbitrarily long to pass any boundary. From the divergent mean RT it follows that the relaxation time of the diffusion process diverges. Together, these results establish that $\mathcal{I}$ will take arbitrarily long to explore its entire phase space.  Consequently, for any finite measurement time $\tau_m$, $\mathcal{I}$ is a non-ergodic process with unequal time and ensemble averages. A corollary to these results is that ergodicity is recovered in the limit $\tau_m\to\infty$, which allows $\mathcal{I}$ to explore and distribute its occupation time across its entire space.

Before concluding, we briefly comment on the practical implications of exploiting WEB in optical sensing.  The averages we have analyzed consume equal amount of energy, and in that sense their comparison is fair. However, the hardware involved is not the same. For the time average $\overline{\mathcal{I}}$ we assumed a single cavity driven by a single laser with power $A^2$. For the ensemble average $\langle\mathcal{I}\rangle_t$ with equal time as $\overline{\mathcal{I}}$, we assumed that the power $A^2$ of one laser is distributed equally among $m$ cavities. The distribution of power requires $m-1$ beam splitters, and $m$ photodetectors would be required for measurements in parallel.    Then, for the ensemble average $\langle\mathcal{I}\rangle_P$ with equal power as $\overline{\mathcal{I}}$, we  would need  $m$ photodetectors and either $m$ lasers with power $A^2$ or a single laser with power $mA^2$ combined with $m-1$ beam splitters.  We recognize that not every application justifies these additional hardware requirements, but some do. Consider, for example, sensors aimed at detecting catastrophic failures or short events with important consequences. In those cases, the benefits of ensemble averaging we have found  can be worth the additional hardware investment.  We also note that, if Markovian dynamics can be rightly assumed, the $m$ resonators are not strictly required to perform ensemble averaging. One could partition trajectories as we have done in Fig.~\ref{fig_2} with our experimental data from one cavity, and perform ensemble averaging on those parts. Such an approach is much simpler to implement, but it will not benefit from the parallelism of the $m$ resonators to improve speed.
	
To summarize, we reported signatures of WEB in the time-integrated intensity $\mathcal{I}$ transmitted by a resonant optical sensor. We compared the sensing precision attained for three distinct averages of $\mathcal{I}$, all as a function of the measurement time.  One is the time average, and the other two are ensemble averages with either equal power or equal measurement time as the time average. We found that the `equal power' ensemble average outperforms the other two averages, by orders of magnitude, for short measurements times. However, this advantage disappears as the measurement time increases and ergodicity emerges. While we fully focused on measurements of $\mathcal{I}$, we expect our results to hold when sensing with other time-integrated observables obeying the arcsine laws. Examples of such observables include the time-integrated real and imaginary parts of $\alpha$~\cite{comp}, which can be measured using balanced homodyne detection~\cite{Lvovsky}. Similarly, while the specific sensor we studied in this manuscript is a single-mode linear optical cavity with additive noise, we expect our results to hold for different systems and conditions. For instance,  for a nonlinear cavity and for a linear cavity influenced by multiplicative noise, our companion letter shows that the arcsine laws hold~\cite{comp}. Consequently, we also expect WEB in those cases. Another important extension of our results is to sensors based on two or more coupled resonators, which have stimulated an important debate about the effects of noise on sensing~\cite{Langbein18, Lau18, Mortensen18, Wiersig20, Duggan22}.  We expect sensing with the time-integrated fields or intensities of those systems to display WEB for the following reason. The intensity is the sum of the squared real and imaginary parts of $\alpha$. Thus, while our system is single mode in the optics parlance, we are already dealing with the sum of two stochastic variables. Furthermore, Equation~\ref{eq1} is mathematically equivalent to models of two-dimensional Brownian motion under the influence of both conservative and non-conservative forces~\cite{Peters23}. That the arcsine laws hold in such general conditions, in two dimensions, strongly suggests that our findings about WEB in optical sensing should hold for higher dimensional systems involving two or more resonators.

Finally, we highlight two perspectives of our work. At the system level, we think that resonators with memory in their nonlinear response~\cite{Geng20, Peters21} offer interesting opportunities to explore WEB beyond the domain of the arcsine laws. The arcsine laws were derived assuming memoryless Markovian dynamics, and are expected to be modified or break down in non-Markovian systems. However, to what extent the conclusions of this manuscript are modified in the presence of non-Markovian dynamics is an open question. The second perspective we offer is at the application level. While our manuscript focused entirely on sensing,  we expect our findings to be relevant to other information processing tasks where time-integrated observables are measured. For instance, nonlinear optical microcavities  are under intense investigation for their potential to solve computational or optimization problems in a probabilistic way~\cite{Liew19, Opala19, Stroev23}.  In those cases as well, understanding the differences between time and ensemble averaging is crucial to make the most precise computation given an energy budget. \\

\section*{Acknowledgments}
\noindent This work is part of the research programme of the Netherlands Organisation for Scientific Research (NWO). We thank Kevin Peters for discussions.  S.R.K.R. acknowledges an ERC Starting Grant with project number 852694. \\


\begin{thebibliography}{62}%
\makeatletter
\providecommand \@ifxundefined [1]{%
 \@ifx{#1\undefined}
}%
\providecommand \@ifnum [1]{%
 \ifnum #1\expandafter \@firstoftwo
 \else \expandafter \@secondoftwo
 \fi
}%
\providecommand \@ifx [1]{%
 \ifx #1\expandafter \@firstoftwo
 \else \expandafter \@secondoftwo
 \fi
}%
\providecommand \natexlab [1]{#1}%
\providecommand \enquote  [1]{``#1''}%
\providecommand \bibnamefont  [1]{#1}%
\providecommand \bibfnamefont [1]{#1}%
\providecommand \citenamefont [1]{#1}%
\providecommand \href@noop [0]{\@secondoftwo}%
\providecommand \href [0]{\begingroup \@sanitize@url \@href}%
\providecommand \@href[1]{\@@startlink{#1}\@@href}%
\providecommand \@@href[1]{\endgroup#1\@@endlink}%
\providecommand \@sanitize@url [0]{\catcode `\\12\catcode `\$12\catcode
  `\&12\catcode `\#12\catcode `\^12\catcode `\_12\catcode `\%12\relax}%
\providecommand \@@startlink[1]{}%
\providecommand \@@endlink[0]{}%
\providecommand \url  [0]{\begingroup\@sanitize@url \@url }%
\providecommand \@url [1]{\endgroup\@href {#1}{\urlprefix }}%
\providecommand \urlprefix  [0]{URL }%
\providecommand \Eprint [0]{\href }%
\providecommand \doibase [0]{https://doi.org/}%
\providecommand \selectlanguage [0]{\@gobble}%
\providecommand \bibinfo  [0]{\@secondoftwo}%
\providecommand \bibfield  [0]{\@secondoftwo}%
\providecommand \translation [1]{[#1]}%
\providecommand \BibitemOpen [0]{}%
\providecommand \bibitemStop [0]{}%
\providecommand \bibitemNoStop [0]{.\EOS\space}%
\providecommand \EOS [0]{\spacefactor3000\relax}%
\providecommand \BibitemShut  [1]{\csname bibitem#1\endcsname}%
\let\auto@bib@innerbib\@empty
\bibitem [{\citenamefont {Zhu}\ \emph {et~al.}(2010)\citenamefont {Zhu},
  \citenamefont {Ozdemir}, \citenamefont {Xiao}, \citenamefont {Li},
  \citenamefont {He}, \citenamefont {Chen},\ and\ \citenamefont
  {Yang}}]{Zhu10}%
  \BibitemOpen
  \bibfield  {author} {\bibinfo {author} {\bibfnamefont {J.}~\bibnamefont
  {Zhu}}, \bibinfo {author} {\bibfnamefont {S.~K.}\ \bibnamefont {Ozdemir}},
  \bibinfo {author} {\bibfnamefont {Y.-F.}\ \bibnamefont {Xiao}}, \bibinfo
  {author} {\bibfnamefont {L.}~\bibnamefont {Li}}, \bibinfo {author}
  {\bibfnamefont {L.}~\bibnamefont {He}}, \bibinfo {author} {\bibfnamefont
  {D.-R.}\ \bibnamefont {Chen}},\ and\ \bibinfo {author} {\bibfnamefont
  {L.}~\bibnamefont {Yang}},\ }\bibfield  {title} {\bibinfo {title} {On-chip
  single nanoparticle detection and sizing by mode splitting in an ultrahigh-q
  microresonator},\ }\href {https://www.nature.com/articles/nphoton.2009.237}
  {\bibfield  {journal} {\bibinfo  {journal} {Nat. Photon.}\ }\textbf {\bibinfo
  {volume} {4}},\ \bibinfo {pages} {46} (\bibinfo {year} {2010})}\BibitemShut
  {NoStop}%
\bibitem [{\citenamefont {Kelkar}\ \emph {et~al.}(2015)\citenamefont {Kelkar},
  \citenamefont {Wang}, \citenamefont {Mart\'{\i}n-Cano}, \citenamefont
  {Hoffmann}, \citenamefont {Christiansen}, \citenamefont {G\"otzinger},\ and\
  \citenamefont {Sandoghdar}}]{Kelkar15}%
  \BibitemOpen
  \bibfield  {author} {\bibinfo {author} {\bibfnamefont {H.}~\bibnamefont
  {Kelkar}}, \bibinfo {author} {\bibfnamefont {D.}~\bibnamefont {Wang}},
  \bibinfo {author} {\bibfnamefont {D.}~\bibnamefont {Mart\'{\i}n-Cano}},
  \bibinfo {author} {\bibfnamefont {B.}~\bibnamefont {Hoffmann}}, \bibinfo
  {author} {\bibfnamefont {S.}~\bibnamefont {Christiansen}}, \bibinfo {author}
  {\bibfnamefont {S.}~\bibnamefont {G\"otzinger}},\ and\ \bibinfo {author}
  {\bibfnamefont {V.}~\bibnamefont {Sandoghdar}},\ }\bibfield  {title}
  {\bibinfo {title} {Sensing nanoparticles with a cantilever-based scannable
  optical cavity of low finesse and sub-${\ensuremath{\lambda}}^{3}$ volume},\
  }\href {https://doi.org/10.1103/PhysRevApplied.4.054010} {\bibfield
  {journal} {\bibinfo  {journal} {Phys. Rev. Appl.}\ }\textbf {\bibinfo
  {volume} {4}},\ \bibinfo {pages} {054010} (\bibinfo {year}
  {2015})}\BibitemShut {NoStop}%
\bibitem [{\citenamefont {Zhi}\ \emph {et~al.}(2017)\citenamefont {Zhi},
  \citenamefont {Yu}, \citenamefont {Gong}, \citenamefont {Yang},\ and\
  \citenamefont {Xiao}}]{Zhi17}%
  \BibitemOpen
  \bibfield  {author} {\bibinfo {author} {\bibfnamefont {Y.}~\bibnamefont
  {Zhi}}, \bibinfo {author} {\bibfnamefont {X.-C.}\ \bibnamefont {Yu}},
  \bibinfo {author} {\bibfnamefont {Q.}~\bibnamefont {Gong}}, \bibinfo {author}
  {\bibfnamefont {L.}~\bibnamefont {Yang}},\ and\ \bibinfo {author}
  {\bibfnamefont {Y.-F.}\ \bibnamefont {Xiao}},\ }\bibfield  {title} {\bibinfo
  {title} {Single nanoparticle detection using optical microcavities},\ }\href
  {https://doi.org/10.1002/adma.201604920} {\bibfield  {journal} {\bibinfo
  {journal} {Adv. Mater.}\ }\textbf {\bibinfo {volume} {29}},\ \bibinfo {pages}
  {1604920} (\bibinfo {year} {2017})}\BibitemShut {NoStop}%
\bibitem [{\citenamefont {Vollmer}\ and\ \citenamefont
  {Arnold}(2008)}]{Vollmer08}%
  \BibitemOpen
  \bibfield  {author} {\bibinfo {author} {\bibfnamefont {F.}~\bibnamefont
  {Vollmer}}\ and\ \bibinfo {author} {\bibfnamefont {S.}~\bibnamefont
  {Arnold}},\ }\bibfield  {title} {\bibinfo {title} {Whispering-gallery-mode
  biosensing: label-free detection down to single molecules},\ }\href
  {https://www.nature.com/articles/nmeth.1221} {\bibfield  {journal} {\bibinfo
  {journal} {Nat. Methods}\ }\textbf {\bibinfo {volume} {5}},\ \bibinfo {pages}
  {591} (\bibinfo {year} {2008})}\BibitemShut {NoStop}%
\bibitem [{\citenamefont {Subramanian}\ \emph {et~al.}(2018)\citenamefont
  {Subramanian}, \citenamefont {Wu}, \citenamefont {Constant}, \citenamefont
  {Xavier},\ and\ \citenamefont {Vollmer}}]{Subramanian18}%
  \BibitemOpen
  \bibfield  {author} {\bibinfo {author} {\bibfnamefont {S.}~\bibnamefont
  {Subramanian}}, \bibinfo {author} {\bibfnamefont {H.-Y.}\ \bibnamefont {Wu}},
  \bibinfo {author} {\bibfnamefont {T.}~\bibnamefont {Constant}}, \bibinfo
  {author} {\bibfnamefont {J.}~\bibnamefont {Xavier}},\ and\ \bibinfo {author}
  {\bibfnamefont {F.}~\bibnamefont {Vollmer}},\ }\bibfield  {title} {\bibinfo
  {title} {Label-free optical single-molecule micro- and nanosensors},\ }\href
  {https://doi.org/https://doi.org/10.1002/adma.201801246} {\bibfield
  {journal} {\bibinfo  {journal} {Adv. Mater.}\ }\textbf {\bibinfo {volume}
  {30}},\ \bibinfo {pages} {1801246} (\bibinfo {year} {2018})}\BibitemShut
  {NoStop}%
\bibitem [{\citenamefont {Feis}\ \emph {et~al.}(2020)\citenamefont {Feis},
  \citenamefont {Beutel}, \citenamefont {K\"opfler}, \citenamefont
  {Garcia-Santiago}, \citenamefont {Rockstuhl}, \citenamefont {Wegener},\ and\
  \citenamefont {Fernandez-Corbaton}}]{Feis20}%
  \BibitemOpen
  \bibfield  {author} {\bibinfo {author} {\bibfnamefont {J.}~\bibnamefont
  {Feis}}, \bibinfo {author} {\bibfnamefont {D.}~\bibnamefont {Beutel}},
  \bibinfo {author} {\bibfnamefont {J.}~\bibnamefont {K\"opfler}}, \bibinfo
  {author} {\bibfnamefont {X.}~\bibnamefont {Garcia-Santiago}}, \bibinfo
  {author} {\bibfnamefont {C.}~\bibnamefont {Rockstuhl}}, \bibinfo {author}
  {\bibfnamefont {M.}~\bibnamefont {Wegener}},\ and\ \bibinfo {author}
  {\bibfnamefont {I.}~\bibnamefont {Fernandez-Corbaton}},\ }\bibfield  {title}
  {\bibinfo {title} {Helicity-preserving optical cavity modes for enhanced
  sensing of chiral molecules},\ }\href
  {https://doi.org/10.1103/PhysRevLett.124.033201} {\bibfield  {journal}
  {\bibinfo  {journal} {Phys. Rev. Lett.}\ }\textbf {\bibinfo {volume} {124}},\
  \bibinfo {pages} {033201} (\bibinfo {year} {2020})}\BibitemShut {NoStop}%
\bibitem [{\citenamefont {Guha}\ \emph {et~al.}(2020)\citenamefont {Guha},
  \citenamefont {Allain}, \citenamefont {Lema\^{\i}tre}, \citenamefont {Leo},\
  and\ \citenamefont {Favero}}]{Guha20}%
  \BibitemOpen
  \bibfield  {author} {\bibinfo {author} {\bibfnamefont {B.}~\bibnamefont
  {Guha}}, \bibinfo {author} {\bibfnamefont {P.~E.}\ \bibnamefont {Allain}},
  \bibinfo {author} {\bibfnamefont {A.}~\bibnamefont {Lema\^{\i}tre}}, \bibinfo
  {author} {\bibfnamefont {G.}~\bibnamefont {Leo}},\ and\ \bibinfo {author}
  {\bibfnamefont {I.}~\bibnamefont {Favero}},\ }\bibfield  {title} {\bibinfo
  {title} {Force sensing with an optomechanical self-oscillator},\ }\href
  {https://doi.org/10.1103/PhysRevApplied.14.024079} {\bibfield  {journal}
  {\bibinfo  {journal} {Phys. Rev. Appl.}\ }\textbf {\bibinfo {volume} {14}},\
  \bibinfo {pages} {024079} (\bibinfo {year} {2020})}\BibitemShut {NoStop}%
\bibitem [{\citenamefont {Oksenberg}\ \emph {et~al.}(2021)\citenamefont
  {Oksenberg}, \citenamefont {Shlesinger}, \citenamefont {Xomalis},
  \citenamefont {Baldi}, \citenamefont {Baumberg}, \citenamefont {Koenderink},\
  and\ \citenamefont {Garnett}}]{Oksenberg21}%
  \BibitemOpen
  \bibfield  {author} {\bibinfo {author} {\bibfnamefont {E.}~\bibnamefont
  {Oksenberg}}, \bibinfo {author} {\bibfnamefont {I.}~\bibnamefont
  {Shlesinger}}, \bibinfo {author} {\bibfnamefont {A.}~\bibnamefont {Xomalis}},
  \bibinfo {author} {\bibfnamefont {A.}~\bibnamefont {Baldi}}, \bibinfo
  {author} {\bibfnamefont {J.~J.}\ \bibnamefont {Baumberg}}, \bibinfo {author}
  {\bibfnamefont {A.~F.}\ \bibnamefont {Koenderink}},\ and\ \bibinfo {author}
  {\bibfnamefont {E.~C.}\ \bibnamefont {Garnett}},\ }\bibfield  {title}
  {\bibinfo {title} {Energy-resolved plasmonic chemistry in individual
  nanoreactors},\ }\href {https://www.nature.com/articles/s41565-021-00973-6}
  {\bibfield  {journal} {\bibinfo  {journal} {Nat. Nanotechnol.}\ }\textbf
  {\bibinfo {volume} {16}},\ \bibinfo {pages} {1378} (\bibinfo {year}
  {2021})}\BibitemShut {NoStop}%
\bibitem [{\citenamefont {Yang}\ \emph {et~al.}(2020)\citenamefont {Yang},
  \citenamefont {Skarda}, \citenamefont {Cotrufo}, \citenamefont {Dutt},
  \citenamefont {Ahn}, \citenamefont {Sawaby}, \citenamefont {Vercruysse},
  \citenamefont {Arbabian}, \citenamefont {Fan}, \citenamefont {Al{\`u}} \emph
  {et~al.}}]{Yang20}%
  \BibitemOpen
  \bibfield  {author} {\bibinfo {author} {\bibfnamefont {K.~Y.}\ \bibnamefont
  {Yang}}, \bibinfo {author} {\bibfnamefont {J.}~\bibnamefont {Skarda}},
  \bibinfo {author} {\bibfnamefont {M.}~\bibnamefont {Cotrufo}}, \bibinfo
  {author} {\bibfnamefont {A.}~\bibnamefont {Dutt}}, \bibinfo {author}
  {\bibfnamefont {G.~H.}\ \bibnamefont {Ahn}}, \bibinfo {author} {\bibfnamefont
  {M.}~\bibnamefont {Sawaby}}, \bibinfo {author} {\bibfnamefont
  {D.}~\bibnamefont {Vercruysse}}, \bibinfo {author} {\bibfnamefont
  {A.}~\bibnamefont {Arbabian}}, \bibinfo {author} {\bibfnamefont
  {S.}~\bibnamefont {Fan}}, \bibinfo {author} {\bibfnamefont {A.}~\bibnamefont
  {Al{\`u}}}, \emph {et~al.},\ }\bibfield  {title} {\bibinfo {title}
  {Inverse-designed non-reciprocal pulse router for chip-based lidar},\ }\href
  {https://www.nature.com/articles/s41566-020-0606-0} {\bibfield  {journal}
  {\bibinfo  {journal} {Nat. Photonics}\ }\textbf {\bibinfo {volume} {14}},\
  \bibinfo {pages} {369} (\bibinfo {year} {2020})}\BibitemShut {NoStop}%
\bibitem [{\citenamefont {Lewis-Swan}\ \emph {et~al.}(2020)\citenamefont
  {Lewis-Swan}, \citenamefont {Barberena}, \citenamefont {Muniz}, \citenamefont
  {Cline}, \citenamefont {Young}, \citenamefont {Thompson},\ and\ \citenamefont
  {Rey}}]{Lewis20}%
  \BibitemOpen
  \bibfield  {author} {\bibinfo {author} {\bibfnamefont {R.~J.}\ \bibnamefont
  {Lewis-Swan}}, \bibinfo {author} {\bibfnamefont {D.}~\bibnamefont
  {Barberena}}, \bibinfo {author} {\bibfnamefont {J.~A.}\ \bibnamefont
  {Muniz}}, \bibinfo {author} {\bibfnamefont {J.~R.~K.}\ \bibnamefont {Cline}},
  \bibinfo {author} {\bibfnamefont {D.}~\bibnamefont {Young}}, \bibinfo
  {author} {\bibfnamefont {J.~K.}\ \bibnamefont {Thompson}},\ and\ \bibinfo
  {author} {\bibfnamefont {A.~M.}\ \bibnamefont {Rey}},\ }\bibfield  {title}
  {\bibinfo {title} {Protocol for precise field sensing in the optical domain
  with cold atoms in a cavity},\ }\href
  {https://doi.org/10.1103/PhysRevLett.124.193602} {\bibfield  {journal}
  {\bibinfo  {journal} {Phys. Rev. Lett.}\ }\textbf {\bibinfo {volume} {124}},\
  \bibinfo {pages} {193602} (\bibinfo {year} {2020})}\BibitemShut {NoStop}%
\bibitem [{\citenamefont {Abbott}\ \emph {et~al.}(2016)\citenamefont {Abbott}
  \emph {et~al.}}]{Abbott16}%
  \BibitemOpen
  \bibfield  {author} {\bibinfo {author} {\bibfnamefont {B.~P.}\ \bibnamefont
  {Abbott}} \emph {et~al.} (\bibinfo {collaboration} {LIGO Scientific
  Collaboration and Virgo Collaboration}),\ }\bibfield  {title} {\bibinfo
  {title} {Observation of gravitational waves from a binary black hole
  merger},\ }\href {https://doi.org/10.1103/PhysRevLett.116.061102} {\bibfield
  {journal} {\bibinfo  {journal} {Phys. Rev. Lett.}\ }\textbf {\bibinfo
  {volume} {116}},\ \bibinfo {pages} {061102} (\bibinfo {year}
  {2016})}\BibitemShut {NoStop}%
\bibitem [{\citenamefont {Lebowitz}\ and\ \citenamefont
  {Penrose}(1973)}]{Lebowitz}%
  \BibitemOpen
  \bibfield  {author} {\bibinfo {author} {\bibfnamefont {J.~L.}\ \bibnamefont
  {Lebowitz}}\ and\ \bibinfo {author} {\bibfnamefont {O.}~\bibnamefont
  {Penrose}},\ }\bibfield  {title} {\bibinfo {title} {{Modern ergodic
  theory}},\ }\href {https://doi.org/10.1063/1.3127948} {\bibfield  {journal}
  {\bibinfo  {journal} {Phys. Today}\ }\textbf {\bibinfo {volume} {26}},\
  \bibinfo {pages} {23} (\bibinfo {year} {1973})}\BibitemShut {NoStop}%
\bibitem [{\citenamefont {Moore}(2015)}]{Moore15}%
  \BibitemOpen
  \bibfield  {author} {\bibinfo {author} {\bibfnamefont {C.~C.}\ \bibnamefont
  {Moore}},\ }\bibfield  {title} {\bibinfo {title} {Ergodic theorem, ergodic
  theory, and statistical mechanics},\ }\href
  {https://doi.org/10.1073/pnas.1421798112} {\bibfield  {journal} {\bibinfo
  {journal} {Proc. Natl. Acad. Sci.}\ }\textbf {\bibinfo {volume} {112}},\
  \bibinfo {pages} {1907} (\bibinfo {year} {2015})}\BibitemShut {NoStop}%
\bibitem [{\citenamefont {Bouchaud}(1992)}]{Bouchaud1992}%
  \BibitemOpen
  \bibfield  {author} {\bibinfo {author} {\bibfnamefont {J.~P.}\ \bibnamefont
  {Bouchaud}},\ }\bibfield  {title} {\bibinfo {title} {Weak ergodicity breaking
  and aging in disordered systems},\ }\href
  {https://doi.org/10.1051/jp1:1992238} {\bibfield  {journal} {\bibinfo
  {journal} {J. phys., I}\ }\textbf {\bibinfo {volume} {2}},\ \bibinfo {pages}
  {1705} (\bibinfo {year} {1992})}\BibitemShut {NoStop}%
\bibitem [{\citenamefont {Bernaschi}\ \emph {et~al.}(2020)\citenamefont
  {Bernaschi}, \citenamefont {Billoire}, \citenamefont {Maiorano},
  \citenamefont {Parisi},\ and\ \citenamefont {Ricci-Tersenghi}}]{Bernaschi20}%
  \BibitemOpen
  \bibfield  {author} {\bibinfo {author} {\bibfnamefont {M.}~\bibnamefont
  {Bernaschi}}, \bibinfo {author} {\bibfnamefont {A.}~\bibnamefont {Billoire}},
  \bibinfo {author} {\bibfnamefont {A.}~\bibnamefont {Maiorano}}, \bibinfo
  {author} {\bibfnamefont {G.}~\bibnamefont {Parisi}},\ and\ \bibinfo {author}
  {\bibfnamefont {F.}~\bibnamefont {Ricci-Tersenghi}},\ }\bibfield  {title}
  {\bibinfo {title} {Strong ergodicity breaking in aging of mean-field spin
  glasses},\ }\href {https://www.pnas.org/doi/abs/10.1073/pnas.1910936117}
  {\bibfield  {journal} {\bibinfo  {journal} {Proc. Natl. Acad. Sci.}\ }\textbf
  {\bibinfo {volume} {117}},\ \bibinfo {pages} {17522} (\bibinfo {year}
  {2020})}\BibitemShut {NoStop}%
\bibitem [{\citenamefont {Cugliandolo}\ and\ \citenamefont
  {Kurchan}(1995)}]{Cugliandolo95}%
  \BibitemOpen
  \bibfield  {author} {\bibinfo {author} {\bibfnamefont {L.~F.}\ \bibnamefont
  {Cugliandolo}}\ and\ \bibinfo {author} {\bibfnamefont {J.}~\bibnamefont
  {Kurchan}},\ }\bibfield  {title} {\bibinfo {title} {Weak ergodicity breaking
  in mean-field spin-glass models},\ }\href
  {https://doi.org/10.1080/01418639508238541} {\bibfield  {journal} {\bibinfo
  {journal} {Philos. mag., B}\ }\textbf {\bibinfo {volume} {71}},\ \bibinfo
  {pages} {501} (\bibinfo {year} {1995})}\BibitemShut {NoStop}%
\bibitem [{\citenamefont {Folena}\ and\ \citenamefont
  {Zamponi}(2023)}]{Folema23}%
  \BibitemOpen
  \bibfield  {author} {\bibinfo {author} {\bibfnamefont {G.}~\bibnamefont
  {Folena}}\ and\ \bibinfo {author} {\bibfnamefont {F.}~\bibnamefont
  {Zamponi}},\ }\bibfield  {title} {\bibinfo {title} {{On weak ergodicity
  breaking in mean-field spin glasses}},\ }\href
  {https://doi.org/10.21468/SciPostPhys.15.3.109} {\bibfield  {journal}
  {\bibinfo  {journal} {SciPost Phys.}\ }\textbf {\bibinfo {volume} {15}},\
  \bibinfo {pages} {109} (\bibinfo {year} {2023})}\BibitemShut {NoStop}%
\bibitem [{\citenamefont {Polkovnikov}\ \emph {et~al.}(2011)\citenamefont
  {Polkovnikov}, \citenamefont {Sengupta}, \citenamefont {Silva},\ and\
  \citenamefont {Vengalattore}}]{Polkovnikov11}%
  \BibitemOpen
  \bibfield  {author} {\bibinfo {author} {\bibfnamefont {A.}~\bibnamefont
  {Polkovnikov}}, \bibinfo {author} {\bibfnamefont {K.}~\bibnamefont
  {Sengupta}}, \bibinfo {author} {\bibfnamefont {A.}~\bibnamefont {Silva}},\
  and\ \bibinfo {author} {\bibfnamefont {M.}~\bibnamefont {Vengalattore}},\
  }\bibfield  {title} {\bibinfo {title} {Colloquium: Nonequilibrium dynamics of
  closed interacting quantum systems},\ }\href
  {https://doi.org/10.1103/RevModPhys.83.863} {\bibfield  {journal} {\bibinfo
  {journal} {Rev. Mod. Phys.}\ }\textbf {\bibinfo {volume} {83}},\ \bibinfo
  {pages} {863} (\bibinfo {year} {2011})}\BibitemShut {NoStop}%
\bibitem [{\citenamefont {Russomanno}\ \emph {et~al.}(2022)\citenamefont
  {Russomanno}, \citenamefont {Fava},\ and\ \citenamefont
  {Fazio}}]{Russomanno22}%
  \BibitemOpen
  \bibfield  {author} {\bibinfo {author} {\bibfnamefont {A.}~\bibnamefont
  {Russomanno}}, \bibinfo {author} {\bibfnamefont {M.}~\bibnamefont {Fava}},\
  and\ \bibinfo {author} {\bibfnamefont {R.}~\bibnamefont {Fazio}},\ }\bibfield
   {title} {\bibinfo {title} {Weak ergodicity breaking in josephson-junction
  arrays},\ }\href {https://doi.org/10.1103/PhysRevB.106.035123} {\bibfield
  {journal} {\bibinfo  {journal} {Phys. Rev. B}\ }\textbf {\bibinfo {volume}
  {106}},\ \bibinfo {pages} {035123} (\bibinfo {year} {2022})}\BibitemShut
  {NoStop}%
\bibitem [{\citenamefont {Surace}\ \emph {et~al.}(2020)\citenamefont {Surace},
  \citenamefont {Giudici},\ and\ \citenamefont {Dalmonte}}]{Surace20}%
  \BibitemOpen
  \bibfield  {author} {\bibinfo {author} {\bibfnamefont {F.~M.}\ \bibnamefont
  {Surace}}, \bibinfo {author} {\bibfnamefont {G.}~\bibnamefont {Giudici}},\
  and\ \bibinfo {author} {\bibfnamefont {M.}~\bibnamefont {Dalmonte}},\
  }\bibfield  {title} {\bibinfo {title} {Weak-ergodicity-breaking via lattice
  supersymmetry},\ }\href {https://doi.org/10.22331/q-2020-10-07-339}
  {\bibfield  {journal} {\bibinfo  {journal} {{Quantum}}\ }\textbf {\bibinfo
  {volume} {4}},\ \bibinfo {pages} {339} (\bibinfo {year} {2020})}\BibitemShut
  {NoStop}%
\bibitem [{\citenamefont {Turner}\ \emph {et~al.}(2018)\citenamefont {Turner},
  \citenamefont {Michailidis}, \citenamefont {Abanin}, \citenamefont {Serbyn},\
  and\ \citenamefont {Papi{\'{c}}}}]{Turner18}%
  \BibitemOpen
  \bibfield  {author} {\bibinfo {author} {\bibfnamefont {C.~J.}\ \bibnamefont
  {Turner}}, \bibinfo {author} {\bibfnamefont {A.~A.}\ \bibnamefont
  {Michailidis}}, \bibinfo {author} {\bibfnamefont {D.~A.}\ \bibnamefont
  {Abanin}}, \bibinfo {author} {\bibfnamefont {M.}~\bibnamefont {Serbyn}},\
  and\ \bibinfo {author} {\bibfnamefont {Z.}~\bibnamefont {Papi{\'{c}}}},\
  }\bibfield  {title} {\bibinfo {title} {Weak ergodicity breaking from quantum
  many-body scars},\ }\href {https://doi.org/10.1038/s41567-018-0137-5}
  {\bibfield  {journal} {\bibinfo  {journal} {Nat. Phys.}\ }\textbf {\bibinfo
  {volume} {14}},\ \bibinfo {pages} {745} (\bibinfo {year} {2018})}\BibitemShut
  {NoStop}%
\bibitem [{\citenamefont {Schecter}\ and\ \citenamefont
  {Iadecola}(2019)}]{Schecter19}%
  \BibitemOpen
  \bibfield  {author} {\bibinfo {author} {\bibfnamefont {M.}~\bibnamefont
  {Schecter}}\ and\ \bibinfo {author} {\bibfnamefont {T.}~\bibnamefont
  {Iadecola}},\ }\bibfield  {title} {\bibinfo {title} {Weak ergodicity breaking
  and quantum many-body scars in spin-1 $xy$ magnets},\ }\href
  {https://doi.org/10.1103/PhysRevLett.123.147201} {\bibfield  {journal}
  {\bibinfo  {journal} {Phys. Rev. Lett.}\ }\textbf {\bibinfo {volume} {123}},\
  \bibinfo {pages} {147201} (\bibinfo {year} {2019})}\BibitemShut {NoStop}%
\bibitem [{\citenamefont {Serbyn}\ \emph {et~al.}(2021)\citenamefont {Serbyn},
  \citenamefont {Abanin},\ and\ \citenamefont {Papi{\'{c}}}}]{Serbyn21}%
  \BibitemOpen
  \bibfield  {author} {\bibinfo {author} {\bibfnamefont {M.}~\bibnamefont
  {Serbyn}}, \bibinfo {author} {\bibfnamefont {D.~A.}\ \bibnamefont {Abanin}},\
  and\ \bibinfo {author} {\bibfnamefont {Z.}~\bibnamefont {Papi{\'{c}}}},\
  }\bibfield  {title} {\bibinfo {title} {Quantum many-body scars and weak
  breaking of ergodicity},\ }\href {https://doi.org/10.1038/s41567-021-01230-2}
  {\bibfield  {journal} {\bibinfo  {journal} {Nat. Phys.}\ }\textbf {\bibinfo
  {volume} {17}},\ \bibinfo {pages} {675} (\bibinfo {year} {2021})}\BibitemShut
  {NoStop}%
\bibitem [{\citenamefont {Lomholt}\ \emph {et~al.}(2007)\citenamefont
  {Lomholt}, \citenamefont {Zaid},\ and\ \citenamefont {Metzler}}]{Lomholt07}%
  \BibitemOpen
  \bibfield  {author} {\bibinfo {author} {\bibfnamefont {M.~A.}\ \bibnamefont
  {Lomholt}}, \bibinfo {author} {\bibfnamefont {I.~M.}\ \bibnamefont {Zaid}},\
  and\ \bibinfo {author} {\bibfnamefont {R.}~\bibnamefont {Metzler}},\
  }\bibfield  {title} {\bibinfo {title} {Subdiffusion and weak ergodicity
  breaking in the presence of a reactive boundary},\ }\href
  {https://doi.org/10.1103/PhysRevLett.98.200603} {\bibfield  {journal}
  {\bibinfo  {journal} {Phys. Rev. Lett.}\ }\textbf {\bibinfo {volume} {98}},\
  \bibinfo {pages} {200603} (\bibinfo {year} {2007})}\BibitemShut {NoStop}%
\bibitem [{\citenamefont {Jeon}\ \emph {et~al.}(2011)\citenamefont {Jeon},
  \citenamefont {Tejedor}, \citenamefont {Burov}, \citenamefont {Barkai},
  \citenamefont {Selhuber-Unkel}, \citenamefont {Berg-S\o{}rensen},
  \citenamefont {Oddershede},\ and\ \citenamefont {Metzler}}]{Jeon11}%
  \BibitemOpen
  \bibfield  {author} {\bibinfo {author} {\bibfnamefont {J.-H.}\ \bibnamefont
  {Jeon}}, \bibinfo {author} {\bibfnamefont {V.}~\bibnamefont {Tejedor}},
  \bibinfo {author} {\bibfnamefont {S.}~\bibnamefont {Burov}}, \bibinfo
  {author} {\bibfnamefont {E.}~\bibnamefont {Barkai}}, \bibinfo {author}
  {\bibfnamefont {C.}~\bibnamefont {Selhuber-Unkel}}, \bibinfo {author}
  {\bibfnamefont {K.}~\bibnamefont {Berg-S\o{}rensen}}, \bibinfo {author}
  {\bibfnamefont {L.}~\bibnamefont {Oddershede}},\ and\ \bibinfo {author}
  {\bibfnamefont {R.}~\bibnamefont {Metzler}},\ }\bibfield  {title} {\bibinfo
  {title} {In vivo anomalous diffusion and weak ergodicity breaking of lipid
  granules},\ }\href {https://doi.org/10.1103/PhysRevLett.106.048103}
  {\bibfield  {journal} {\bibinfo  {journal} {Phys. Rev. Lett.}\ }\textbf
  {\bibinfo {volume} {106}},\ \bibinfo {pages} {048103} (\bibinfo {year}
  {2011})}\BibitemShut {NoStop}%
\bibitem [{\citenamefont {Manzo}\ \emph {et~al.}(2015)\citenamefont {Manzo},
  \citenamefont {Torreno-Pina}, \citenamefont {Massignan}, \citenamefont
  {Lapeyre}, \citenamefont {Lewenstein},\ and\ \citenamefont
  {Garcia~Parajo}}]{Manzo15}%
  \BibitemOpen
  \bibfield  {author} {\bibinfo {author} {\bibfnamefont {C.}~\bibnamefont
  {Manzo}}, \bibinfo {author} {\bibfnamefont {J.~A.}\ \bibnamefont
  {Torreno-Pina}}, \bibinfo {author} {\bibfnamefont {P.}~\bibnamefont
  {Massignan}}, \bibinfo {author} {\bibfnamefont {G.~J.}\ \bibnamefont
  {Lapeyre}}, \bibinfo {author} {\bibfnamefont {M.}~\bibnamefont
  {Lewenstein}},\ and\ \bibinfo {author} {\bibfnamefont {M.~F.}\ \bibnamefont
  {Garcia~Parajo}},\ }\bibfield  {title} {\bibinfo {title} {Weak ergodicity
  breaking of receptor motion in living cells stemming from random
  diffusivity},\ }\href {https://doi.org/10.1103/PhysRevX.5.011021} {\bibfield
  {journal} {\bibinfo  {journal} {Phys. Rev. X}\ }\textbf {\bibinfo {volume}
  {5}},\ \bibinfo {pages} {011021} (\bibinfo {year} {2015})}\BibitemShut
  {NoStop}%
\bibitem [{\citenamefont {Weron}\ \emph {et~al.}(2017)\citenamefont {Weron},
  \citenamefont {Burnecki}, \citenamefont {Akin}, \citenamefont {Sol{\'e}},
  \citenamefont {Balcerek}, \citenamefont {Tamkun},\ and\ \citenamefont
  {Krapf}}]{Weron2017}%
  \BibitemOpen
  \bibfield  {author} {\bibinfo {author} {\bibfnamefont {A.}~\bibnamefont
  {Weron}}, \bibinfo {author} {\bibfnamefont {K.}~\bibnamefont {Burnecki}},
  \bibinfo {author} {\bibfnamefont {E.~J.}\ \bibnamefont {Akin}}, \bibinfo
  {author} {\bibfnamefont {L.}~\bibnamefont {Sol{\'e}}}, \bibinfo {author}
  {\bibfnamefont {M.}~\bibnamefont {Balcerek}}, \bibinfo {author}
  {\bibfnamefont {M.~M.}\ \bibnamefont {Tamkun}},\ and\ \bibinfo {author}
  {\bibfnamefont {D.}~\bibnamefont {Krapf}},\ }\bibfield  {title} {\bibinfo
  {title} {Ergodicity breaking on the neuronal surface emerges from random
  switching between diffusive states},\ }\href
  {https://doi.org/10.1038/s41598-017-05911-y} {\bibfield  {journal} {\bibinfo
  {journal} {Sci. Rep.}\ }\textbf {\bibinfo {volume} {7}},\ \bibinfo {pages}
  {5404} (\bibinfo {year} {2017})}\BibitemShut {NoStop}%
\bibitem [{\citenamefont {Stojkoski}\ \emph
  {et~al.}(2022{\natexlab{a}})\citenamefont {Stojkoski}, \citenamefont
  {Sandev}, \citenamefont {Kocarev},\ and\ \citenamefont {Pal}}]{Stojkoski22}%
  \BibitemOpen
  \bibfield  {author} {\bibinfo {author} {\bibfnamefont {V.}~\bibnamefont
  {Stojkoski}}, \bibinfo {author} {\bibfnamefont {T.}~\bibnamefont {Sandev}},
  \bibinfo {author} {\bibfnamefont {L.}~\bibnamefont {Kocarev}},\ and\ \bibinfo
  {author} {\bibfnamefont {A.}~\bibnamefont {Pal}},\ }\bibfield  {title}
  {\bibinfo {title} {Autocorrelation functions and ergodicity in diffusion with
  stochastic resetting},\ }\href {https://doi.org/10.1088/1751-8121/ac4ce9}
  {\bibfield  {journal} {\bibinfo  {journal} {J. Phys. A Math. Theor.}\
  }\textbf {\bibinfo {volume} {55}},\ \bibinfo {pages} {104003} (\bibinfo
  {year} {2022}{\natexlab{a}})}\BibitemShut {NoStop}%
\bibitem [{\citenamefont {Singh}(2022)}]{Singh22}%
  \BibitemOpen
  \bibfield  {author} {\bibinfo {author} {\bibfnamefont {P.}~\bibnamefont
  {Singh}},\ }\bibfield  {title} {\bibinfo {title} {Extreme value statistics
  and arcsine laws for heterogeneous diffusion processes},\ }\href
  {https://doi.org/10.1103/PhysRevE.105.024113} {\bibfield  {journal} {\bibinfo
   {journal} {Phys. Rev. E}\ }\textbf {\bibinfo {volume} {105}},\ \bibinfo
  {pages} {024113} (\bibinfo {year} {2022})}\BibitemShut {NoStop}%
\bibitem [{\citenamefont {Cherstvy}\ and\ \citenamefont
  {Metzler}(2015)}]{Cherstvy15}%
  \BibitemOpen
  \bibfield  {author} {\bibinfo {author} {\bibfnamefont {A.~G.}\ \bibnamefont
  {Cherstvy}}\ and\ \bibinfo {author} {\bibfnamefont {R.}~\bibnamefont
  {Metzler}},\ }\bibfield  {title} {\bibinfo {title} {Ergodicity breaking and
  particle spreading in noisy heterogeneous diffusion processes},\ }\href
  {https://doi.org/10.1063/1.4917077} {\bibfield  {journal} {\bibinfo
  {journal} {J. Chem. Phys.}\ }\textbf {\bibinfo {volume} {142}},\ \bibinfo
  {pages} {144105} (\bibinfo {year} {2015})}\BibitemShut {NoStop}%
\bibitem [{\citenamefont {Margolin}\ and\ \citenamefont
  {Barkai}(2005)}]{Margolin05}%
  \BibitemOpen
  \bibfield  {author} {\bibinfo {author} {\bibfnamefont {G.}~\bibnamefont
  {Margolin}}\ and\ \bibinfo {author} {\bibfnamefont {E.}~\bibnamefont
  {Barkai}},\ }\bibfield  {title} {\bibinfo {title} {Nonergodicity of blinking
  nanocrystals and other l\'evy-walk processes},\ }\href
  {https://doi.org/10.1103/PhysRevLett.94.080601} {\bibfield  {journal}
  {\bibinfo  {journal} {Phys. Rev. Lett.}\ }\textbf {\bibinfo {volume} {94}},\
  \bibinfo {pages} {080601} (\bibinfo {year} {2005})}\BibitemShut {NoStop}%
\bibitem [{\citenamefont {Margolin}\ \emph {et~al.}(2006)\citenamefont
  {Margolin}, \citenamefont {Protasenko}, \citenamefont {Kuno},\ and\
  \citenamefont {Barkai}}]{Margolin06}%
  \BibitemOpen
  \bibfield  {author} {\bibinfo {author} {\bibfnamefont {G.}~\bibnamefont
  {Margolin}}, \bibinfo {author} {\bibfnamefont {V.}~\bibnamefont
  {Protasenko}}, \bibinfo {author} {\bibfnamefont {M.}~\bibnamefont {Kuno}},\
  and\ \bibinfo {author} {\bibfnamefont {E.}~\bibnamefont {Barkai}},\
  }\bibfield  {title} {\bibinfo {title} {Photon counting statistics for
  blinking cdse-zns quantum dots: A l\'evy walk process},\ }\href
  {https://doi.org/10.1021/jp061487m} {\bibfield  {journal} {\bibinfo
  {journal} {J. Phys. Chem. B}\ }\textbf {\bibinfo {volume} {110}},\ \bibinfo
  {pages} {19053} (\bibinfo {year} {2006})}\BibitemShut {NoStop}%
\bibitem [{\citenamefont {Burov}\ \emph {et~al.}(2011)\citenamefont {Burov},
  \citenamefont {Jeon}, \citenamefont {Metzler},\ and\ \citenamefont
  {Barkai}}]{Burov11}%
  \BibitemOpen
  \bibfield  {author} {\bibinfo {author} {\bibfnamefont {S.}~\bibnamefont
  {Burov}}, \bibinfo {author} {\bibfnamefont {J.-H.}\ \bibnamefont {Jeon}},
  \bibinfo {author} {\bibfnamefont {R.}~\bibnamefont {Metzler}},\ and\ \bibinfo
  {author} {\bibfnamefont {E.}~\bibnamefont {Barkai}},\ }\bibfield  {title}
  {\bibinfo {title} {Single particle tracking in systems showing anomalous
  diffusion: the role of weak ergodicity breaking},\ }\href
  {https://doi.org/10.1039/C0CP01879A} {\bibfield  {journal} {\bibinfo
  {journal} {Phys. Chem. Chem. Phys.}\ }\textbf {\bibinfo {volume} {13}},\
  \bibinfo {pages} {1800} (\bibinfo {year} {2011})}\BibitemShut {NoStop}%
\bibitem [{\citenamefont {Stojkoski}\ \emph
  {et~al.}(2022{\natexlab{b}})\citenamefont {Stojkoski}, \citenamefont
  {Jolakoski}, \citenamefont {Pal}, \citenamefont {Sandev}, \citenamefont
  {Kocarev},\ and\ \citenamefont {Metzler}}]{Stojkoski2022}%
  \BibitemOpen
  \bibfield  {author} {\bibinfo {author} {\bibfnamefont {V.}~\bibnamefont
  {Stojkoski}}, \bibinfo {author} {\bibfnamefont {P.}~\bibnamefont
  {Jolakoski}}, \bibinfo {author} {\bibfnamefont {A.}~\bibnamefont {Pal}},
  \bibinfo {author} {\bibfnamefont {T.}~\bibnamefont {Sandev}}, \bibinfo
  {author} {\bibfnamefont {L.}~\bibnamefont {Kocarev}},\ and\ \bibinfo {author}
  {\bibfnamefont {R.}~\bibnamefont {Metzler}},\ }\bibfield  {title} {\bibinfo
  {title} {Income inequality and mobility in geometric brownian motion with
  stochastic resetting: theoretical results and empirical evidence of
  non-ergodicity},\ }\href
  {https://royalsocietypublishing.org/doi/10.1098/rsta.2021.0157} {\bibfield
  {journal} {\bibinfo  {journal} {Phil. Trans. R. Soc. A.}\ }\textbf {\bibinfo
  {volume} {380}} (\bibinfo {year} {2022}{\natexlab{b}})}\BibitemShut {NoStop}%
\bibitem [{\citenamefont {Ramesh}\ \emph {et~al.}(2024)\citenamefont {Ramesh},
  \citenamefont {Peters},\ and\ \citenamefont {Rodriguez}}]{comp}%
  \BibitemOpen
  \bibfield  {author} {\bibinfo {author} {\bibfnamefont {V.~G.}\ \bibnamefont
  {Ramesh}}, \bibinfo {author} {\bibfnamefont {K.~J.~H.}\ \bibnamefont
  {Peters}},\ and\ \bibinfo {author} {\bibfnamefont {S.}~\bibnamefont
  {Rodriguez}},\ }\bibfield  {title} {\bibinfo {title} {Arcsine laws of
  light},\ }\href@noop {} {\bibfield  {journal} {\bibinfo  {journal} {Phys.
  Rev. Lett.}\ }\textbf {\bibinfo {volume} {in press}} (\bibinfo {year}
  {2024})}\BibitemShut {NoStop}%
\bibitem [{\citenamefont {Bel}\ and\ \citenamefont {Barkai}(2006)}]{Bel06}%
  \BibitemOpen
  \bibfield  {author} {\bibinfo {author} {\bibfnamefont {G.}~\bibnamefont
  {Bel}}\ and\ \bibinfo {author} {\bibfnamefont {E.}~\bibnamefont {Barkai}},\
  }\bibfield  {title} {\bibinfo {title} {Random walk to a nonergodic
  equilibrium concept},\ }\href {https://doi.org/10.1103/PhysRevE.73.016125}
  {\bibfield  {journal} {\bibinfo  {journal} {Phys. Rev. E}\ }\textbf {\bibinfo
  {volume} {73}},\ \bibinfo {pages} {016125} (\bibinfo {year}
  {2006})}\BibitemShut {NoStop}%
\bibitem [{\citenamefont {Rebenshtok}\ and\ \citenamefont
  {Barkai}(2008)}]{Rebenshtok08}%
  \BibitemOpen
  \bibfield  {author} {\bibinfo {author} {\bibfnamefont {A.}~\bibnamefont
  {Rebenshtok}}\ and\ \bibinfo {author} {\bibfnamefont {E.}~\bibnamefont
  {Barkai}},\ }\bibfield  {title} {\bibinfo {title} {Weakly non-ergodic
  statistical physics},\ }\href {https://doi.org/10.1007/s10955-008-9610-3}
  {\bibfield  {journal} {\bibinfo  {journal} {J. Stat. Phys}\ }\textbf
  {\bibinfo {volume} {133}},\ \bibinfo {pages} {565} (\bibinfo {year}
  {2008})}\BibitemShut {NoStop}%
\bibitem [{\citenamefont {Korabel}\ and\ \citenamefont
  {Barkai}(2013)}]{Korabel13}%
  \BibitemOpen
  \bibfield  {author} {\bibinfo {author} {\bibfnamefont {N.}~\bibnamefont
  {Korabel}}\ and\ \bibinfo {author} {\bibfnamefont {E.}~\bibnamefont
  {Barkai}},\ }\bibfield  {title} {\bibinfo {title} {Distributions of time
  averages for weakly chaotic systems: The role of infinite invariant
  density},\ }\href {https://doi.org/10.1103/PhysRevE.88.032114} {\bibfield
  {journal} {\bibinfo  {journal} {Phys. Rev. E}\ }\textbf {\bibinfo {volume}
  {88}},\ \bibinfo {pages} {032114} (\bibinfo {year} {2013})}\BibitemShut
  {NoStop}%
\bibitem [{\citenamefont {Hunger}\ \emph {et~al.}(2010)\citenamefont {Hunger},
  \citenamefont {Steinmetz}, \citenamefont {Colombe}, \citenamefont {Deutsch},
  \citenamefont {Hänsch},\ and\ \citenamefont {Reichel}}]{Hunger10}%
  \BibitemOpen
  \bibfield  {author} {\bibinfo {author} {\bibfnamefont {D.}~\bibnamefont
  {Hunger}}, \bibinfo {author} {\bibfnamefont {T.}~\bibnamefont {Steinmetz}},
  \bibinfo {author} {\bibfnamefont {Y.}~\bibnamefont {Colombe}}, \bibinfo
  {author} {\bibfnamefont {C.}~\bibnamefont {Deutsch}}, \bibinfo {author}
  {\bibfnamefont {T.~W.}\ \bibnamefont {Hänsch}},\ and\ \bibinfo {author}
  {\bibfnamefont {J.}~\bibnamefont {Reichel}},\ }\bibfield  {title} {\bibinfo
  {title} {A fiber fabry{\textendash}perot cavity with high finesse},\ }\href
  {https://doi.org/10.1088/1367-2630/12/6/065038} {\bibfield  {journal}
  {\bibinfo  {journal} {New J. Phys.}\ }\textbf {\bibinfo {volume} {12}},\
  \bibinfo {pages} {065038} (\bibinfo {year} {2010})}\BibitemShut {NoStop}%
\bibitem [{\citenamefont {Trichet}\ \emph {et~al.}(2014)\citenamefont
  {Trichet}, \citenamefont {Foster}, \citenamefont {Omori}, \citenamefont
  {James}, \citenamefont {Dolan}, \citenamefont {Hughes}, \citenamefont
  {Vallance},\ and\ \citenamefont {Smith}}]{Trichet14}%
  \BibitemOpen
  \bibfield  {author} {\bibinfo {author} {\bibfnamefont {A.~A.~P.}\
  \bibnamefont {Trichet}}, \bibinfo {author} {\bibfnamefont {J.}~\bibnamefont
  {Foster}}, \bibinfo {author} {\bibfnamefont {N.~E.}\ \bibnamefont {Omori}},
  \bibinfo {author} {\bibfnamefont {D.}~\bibnamefont {James}}, \bibinfo
  {author} {\bibfnamefont {P.~R.}\ \bibnamefont {Dolan}}, \bibinfo {author}
  {\bibfnamefont {G.~M.}\ \bibnamefont {Hughes}}, \bibinfo {author}
  {\bibfnamefont {C.}~\bibnamefont {Vallance}},\ and\ \bibinfo {author}
  {\bibfnamefont {J.~M.}\ \bibnamefont {Smith}},\ }\bibfield  {title} {\bibinfo
  {title} {Open-access optical microcavities for lab-on-a-chip refractive index
  sensing},\ }\href {https://doi.org/10.1039/C4LC00817K} {\bibfield  {journal}
  {\bibinfo  {journal} {Lab Chip}\ }\textbf {\bibinfo {volume} {14}},\ \bibinfo
  {pages} {4244} (\bibinfo {year} {2014})}\BibitemShut {NoStop}%
\bibitem [{\citenamefont {Vallance}\ \emph {et~al.}(2016)\citenamefont
  {Vallance}, \citenamefont {Trichet}, \citenamefont {James}, \citenamefont
  {Dolan},\ and\ \citenamefont {Smith}}]{Vallance16}%
  \BibitemOpen
  \bibfield  {author} {\bibinfo {author} {\bibfnamefont {C.}~\bibnamefont
  {Vallance}}, \bibinfo {author} {\bibfnamefont {A.~A.~P.}\ \bibnamefont
  {Trichet}}, \bibinfo {author} {\bibfnamefont {D.}~\bibnamefont {James}},
  \bibinfo {author} {\bibfnamefont {P.~R.}\ \bibnamefont {Dolan}},\ and\
  \bibinfo {author} {\bibfnamefont {J.~M.}\ \bibnamefont {Smith}},\ }\bibfield
  {title} {\bibinfo {title} {Open-access microcavities for chemical sensing},\
  }\href {https://doi.org/10.1088/0957-4484/27/27/274003} {\bibfield  {journal}
  {\bibinfo  {journal} {Nanotechnology}\ }\textbf {\bibinfo {volume} {27}},\
  \bibinfo {pages} {274003} (\bibinfo {year} {2016})}\BibitemShut {NoStop}%
\bibitem [{\citenamefont {Bogaerts}\ \emph {et~al.}(2012)\citenamefont
  {Bogaerts}, \citenamefont {De~Heyn}, \citenamefont {Van~Vaerenbergh},
  \citenamefont {De~Vos}, \citenamefont {Kumar~Selvaraja}, \citenamefont
  {Claes}, \citenamefont {Dumon}, \citenamefont {Bienstman}, \citenamefont
  {Van~Thourhout},\ and\ \citenamefont {Baets}}]{Bogaerts12}%
  \BibitemOpen
  \bibfield  {author} {\bibinfo {author} {\bibfnamefont {W.}~\bibnamefont
  {Bogaerts}}, \bibinfo {author} {\bibfnamefont {P.}~\bibnamefont {De~Heyn}},
  \bibinfo {author} {\bibfnamefont {T.}~\bibnamefont {Van~Vaerenbergh}},
  \bibinfo {author} {\bibfnamefont {K.}~\bibnamefont {De~Vos}}, \bibinfo
  {author} {\bibfnamefont {S.}~\bibnamefont {Kumar~Selvaraja}}, \bibinfo
  {author} {\bibfnamefont {T.}~\bibnamefont {Claes}}, \bibinfo {author}
  {\bibfnamefont {P.}~\bibnamefont {Dumon}}, \bibinfo {author} {\bibfnamefont
  {P.}~\bibnamefont {Bienstman}}, \bibinfo {author} {\bibfnamefont
  {D.}~\bibnamefont {Van~Thourhout}},\ and\ \bibinfo {author} {\bibfnamefont
  {R.}~\bibnamefont {Baets}},\ }\bibfield  {title} {\bibinfo {title} {Silicon
  microring resonators},\ }\href
  {https://doi.org/https://doi.org/10.1002/lpor.201100017} {\bibfield
  {journal} {\bibinfo  {journal} {Laser Photonics Rev.}\ }\textbf {\bibinfo
  {volume} {6}},\ \bibinfo {pages} {47} (\bibinfo {year} {2012})}\BibitemShut
  {NoStop}%
\bibitem [{\citenamefont {Pitruzzello}\ and\ \citenamefont
  {Krauss}(2018)}]{Pitruzzello18}%
  \BibitemOpen
  \bibfield  {author} {\bibinfo {author} {\bibfnamefont {G.}~\bibnamefont
  {Pitruzzello}}\ and\ \bibinfo {author} {\bibfnamefont {T.~F.}\ \bibnamefont
  {Krauss}},\ }\bibfield  {title} {\bibinfo {title} {Photonic crystal
  resonances for sensing and imaging},\ }\href
  {https://doi.org/10.1088/2040-8986/aac75b} {\bibfield  {journal} {\bibinfo
  {journal} {J Opt.}\ }\textbf {\bibinfo {volume} {20}},\ \bibinfo {pages}
  {073004} (\bibinfo {year} {2018})}\BibitemShut {NoStop}%
\bibitem [{\citenamefont {L\'evy}(1940)}]{Levy40}%
  \BibitemOpen
  \bibfield  {author} {\bibinfo {author} {\bibfnamefont {P.}~\bibnamefont
  {L\'evy}},\ }\bibfield  {title} {\bibinfo {title} {Sur certains processus
  stochastiques homog\`enes},\ }\href
  {http://www.numdam.org/item/CM_1940__7__283_0/} {\bibfield  {journal}
  {\bibinfo  {journal} {Compos. Math.}\ }\textbf {\bibinfo {volume} {7}},\
  \bibinfo {pages} {283} (\bibinfo {year} {1940})}\BibitemShut {NoStop}%
\bibitem [{\citenamefont {Barato}\ \emph {et~al.}(2018)\citenamefont {Barato},
  \citenamefont {Rold\'an}, \citenamefont {Mart\'{\i}nez},\ and\ \citenamefont
  {Pigolotti}}]{Barato18}%
  \BibitemOpen
  \bibfield  {author} {\bibinfo {author} {\bibfnamefont {A.~C.}\ \bibnamefont
  {Barato}}, \bibinfo {author} {\bibfnamefont {E.}~\bibnamefont {Rold\'an}},
  \bibinfo {author} {\bibfnamefont {I.~A.}\ \bibnamefont {Mart\'{\i}nez}},\
  and\ \bibinfo {author} {\bibfnamefont {S.}~\bibnamefont {Pigolotti}},\
  }\bibfield  {title} {\bibinfo {title} {Arcsine laws in stochastic
  thermodynamics},\ }\href {https://doi.org/10.1103/PhysRevLett.121.090601}
  {\bibfield  {journal} {\bibinfo  {journal} {Phys. Rev. Lett.}\ }\textbf
  {\bibinfo {volume} {121}},\ \bibinfo {pages} {090601} (\bibinfo {year}
  {2018})}\BibitemShut {NoStop}%
\bibitem [{\citenamefont {Majumdar}\ \emph {et~al.}(2020)\citenamefont
  {Majumdar}, \citenamefont {Pal},\ and\ \citenamefont {Schehr}}]{Majumdar20}%
  \BibitemOpen
  \bibfield  {author} {\bibinfo {author} {\bibfnamefont {S.~N.}\ \bibnamefont
  {Majumdar}}, \bibinfo {author} {\bibfnamefont {A.}~\bibnamefont {Pal}},\ and\
  \bibinfo {author} {\bibfnamefont {G.}~\bibnamefont {Schehr}},\ }\bibfield
  {title} {\bibinfo {title} {Extreme value statistics of correlated random
  variables: A pedagogical review},\ }\href
  {https://doi.org/https://doi.org/10.1016/j.physrep.2019.10.005} {\bibfield
  {journal} {\bibinfo  {journal} {Phys. Rep.}\ }\textbf {\bibinfo {volume}
  {840}},\ \bibinfo {pages} {1} (\bibinfo {year} {2020})}\BibitemShut {NoStop}%
\bibitem [{\citenamefont {Kay}(1998)}]{Kay}%
  \BibitemOpen
  \bibfield  {author} {\bibinfo {author} {\bibfnamefont {S.}~\bibnamefont
  {Kay}},\ }\href@noop {} {\emph {\bibinfo {title} {Fundamentals of Statistical
  Signal Processing: Detection theory}}},\ Prentice Hall Signal Processing
  Series\ (\bibinfo  {publisher} {Prentice-Hall PTR, Hoboken},\ \bibinfo {year}
  {1998})\BibitemShut {NoStop}%
\bibitem [{\citenamefont {Feller}(1957)}]{feller57}%
  \BibitemOpen
  \bibfield  {author} {\bibinfo {author} {\bibfnamefont {W.}~\bibnamefont
  {Feller}},\ }\href@noop {} {\emph {\bibinfo {title} {An Introduction to
  Probability Theory and Its Applications}}},\ \bibinfo {series} {An
  Introduction to Probability Theory and Its Applications}\ No.\ \bibinfo
  {number} {Vols. 1 and 2}\ (\bibinfo  {publisher} {John Wiley \& Sons, New
  York},\ \bibinfo {year} {1957})\BibitemShut {NoStop}%
\bibitem [{\citenamefont {Bhattacharya}\ and\ \citenamefont
  {Waymire}(1990)}]{bhattacharya90}%
  \BibitemOpen
  \bibfield  {author} {\bibinfo {author} {\bibfnamefont {R.}~\bibnamefont
  {Bhattacharya}}\ and\ \bibinfo {author} {\bibfnamefont {E.}~\bibnamefont
  {Waymire}},\ }\href@noop {} {\emph {\bibinfo {title} {Stochastic Processes
  with Applications}}},\ Wiley Series in Probability and Statistics - Applied
  Probability and Statistics Section\ (\bibinfo  {publisher} {John Wiley \&
  Sons, New York},\ \bibinfo {year} {1990})\BibitemShut {NoStop}%
\bibitem [{\citenamefont {Ding}\ and\ \citenamefont {Yang}(1995)}]{Mingzhou95}%
  \BibitemOpen
  \bibfield  {author} {\bibinfo {author} {\bibfnamefont {M.}~\bibnamefont
  {Ding}}\ and\ \bibinfo {author} {\bibfnamefont {W.}~\bibnamefont {Yang}},\
  }\bibfield  {title} {\bibinfo {title} {Distribution of the first return time
  in fractional brownian motion and its application to the study of on-off
  intermittency},\ }\href {https://doi.org/10.1103/PhysRevE.52.207} {\bibfield
  {journal} {\bibinfo  {journal} {Phys. Rev. E}\ }\textbf {\bibinfo {volume}
  {52}},\ \bibinfo {pages} {207} (\bibinfo {year} {1995})}\BibitemShut
  {NoStop}%
\bibitem [{\citenamefont {Lvovsky}\ and\ \citenamefont
  {Raymer}(2009)}]{Lvovsky}%
  \BibitemOpen
  \bibfield  {author} {\bibinfo {author} {\bibfnamefont {A.~I.}\ \bibnamefont
  {Lvovsky}}\ and\ \bibinfo {author} {\bibfnamefont {M.~G.}\ \bibnamefont
  {Raymer}},\ }\bibfield  {title} {\bibinfo {title} {Continuous-variable
  optical quantum-state tomography},\ }\href
  {https://doi.org/10.1103/RevModPhys.81.299} {\bibfield  {journal} {\bibinfo
  {journal} {Rev. Mod. Phys.}\ }\textbf {\bibinfo {volume} {81}},\ \bibinfo
  {pages} {299} (\bibinfo {year} {2009})}\BibitemShut {NoStop}%
\bibitem [{\citenamefont {Langbein}(2018)}]{Langbein18}%
  \BibitemOpen
  \bibfield  {author} {\bibinfo {author} {\bibfnamefont {W.}~\bibnamefont
  {Langbein}},\ }\bibfield  {title} {\bibinfo {title} {No exceptional precision
  of exceptional-point sensors},\ }\href
  {https://doi.org/10.1103/PhysRevA.98.023805} {\bibfield  {journal} {\bibinfo
  {journal} {Phys. Rev. A}\ }\textbf {\bibinfo {volume} {98}},\ \bibinfo
  {pages} {023805} (\bibinfo {year} {2018})}\BibitemShut {NoStop}%
\bibitem [{\citenamefont {Lau}\ and\ \citenamefont {Clerk}(2018)}]{Lau18}%
  \BibitemOpen
  \bibfield  {author} {\bibinfo {author} {\bibfnamefont {H.-K.}\ \bibnamefont
  {Lau}}\ and\ \bibinfo {author} {\bibfnamefont {A.~A.}\ \bibnamefont
  {Clerk}},\ }\bibfield  {title} {\bibinfo {title} {Fundamental limits and
  non-reciprocal approaches in non-{H}ermitian quantum sensing},\ }\href
  {https://www.nature.com/articles/s41467-018-06477-7} {\bibfield  {journal}
  {\bibinfo  {journal} {Nat. Commun.}\ }\textbf {\bibinfo {volume} {9}},\
  \bibinfo {pages} {1} (\bibinfo {year} {2018})}\BibitemShut {NoStop}%
\bibitem [{\citenamefont {Mortensen}\ \emph {et~al.}(2018)\citenamefont
  {Mortensen}, \citenamefont {Gon\c{c}alves}, \citenamefont {Khajavikhan},
  \citenamefont {Christodoulides}, \citenamefont {Tserkezis},\ and\
  \citenamefont {Wolff}}]{Mortensen18}%
  \BibitemOpen
  \bibfield  {author} {\bibinfo {author} {\bibfnamefont {N.~A.}\ \bibnamefont
  {Mortensen}}, \bibinfo {author} {\bibfnamefont {P.~A.~D.}\ \bibnamefont
  {Gon\c{c}alves}}, \bibinfo {author} {\bibfnamefont {M.}~\bibnamefont
  {Khajavikhan}}, \bibinfo {author} {\bibfnamefont {D.~N.}\ \bibnamefont
  {Christodoulides}}, \bibinfo {author} {\bibfnamefont {C.}~\bibnamefont
  {Tserkezis}},\ and\ \bibinfo {author} {\bibfnamefont {C.}~\bibnamefont
  {Wolff}},\ }\bibfield  {title} {\bibinfo {title} {Fluctuations and
  noise-limited sensing near the exceptional point of parity-time-symmetric
  resonator systems},\ }\href {https://doi.org/10.1364/OPTICA.5.001342}
  {\bibfield  {journal} {\bibinfo  {journal} {Optica}\ }\textbf {\bibinfo
  {volume} {5}},\ \bibinfo {pages} {1342} (\bibinfo {year} {2018})}\BibitemShut
  {NoStop}%
\bibitem [{\citenamefont {Wiersig}(2020)}]{Wiersig20}%
  \BibitemOpen
  \bibfield  {author} {\bibinfo {author} {\bibfnamefont {J.}~\bibnamefont
  {Wiersig}},\ }\bibfield  {title} {\bibinfo {title} {Review of exceptional
  point-based sensors},\ }\href {https://doi.org/10.1364/PRJ.396115} {\bibfield
   {journal} {\bibinfo  {journal} {Photon. Res.}\ }\textbf {\bibinfo {volume}
  {8}},\ \bibinfo {pages} {1457} (\bibinfo {year} {2020})}\BibitemShut
  {NoStop}%
\bibitem [{\citenamefont {Duggan}\ \emph {et~al.}(2022)\citenamefont {Duggan},
  \citenamefont {Mann},\ and\ \citenamefont {Alù}}]{Duggan22}%
  \BibitemOpen
  \bibfield  {author} {\bibinfo {author} {\bibfnamefont {R.}~\bibnamefont
  {Duggan}}, \bibinfo {author} {\bibfnamefont {S.~A.}\ \bibnamefont {Mann}},\
  and\ \bibinfo {author} {\bibfnamefont {A.}~\bibnamefont {Alù}},\ }\bibfield
  {title} {\bibinfo {title} {Limitations of sensing at an exceptional point},\
  }\href {https://doi.org/10.1021/acsphotonics.1c01535} {\bibfield  {journal}
  {\bibinfo  {journal} {ACS Photonics}\ }\textbf {\bibinfo {volume} {9}},\
  \bibinfo {pages} {1554} (\bibinfo {year} {2022})}\BibitemShut {NoStop}%
\bibitem [{\citenamefont {Peters}\ \emph {et~al.}(2023)\citenamefont {Peters},
  \citenamefont {Busink}, \citenamefont {Ackermans}, \citenamefont {Cogn\'ee},\
  and\ \citenamefont {Rodriguez}}]{Peters23}%
  \BibitemOpen
  \bibfield  {author} {\bibinfo {author} {\bibfnamefont {K.~J.~H.}\
  \bibnamefont {Peters}}, \bibinfo {author} {\bibfnamefont {J.}~\bibnamefont
  {Busink}}, \bibinfo {author} {\bibfnamefont {P.}~\bibnamefont {Ackermans}},
  \bibinfo {author} {\bibfnamefont {K.~G.}\ \bibnamefont {Cogn\'ee}},\ and\
  \bibinfo {author} {\bibfnamefont {S.~R.~K.}\ \bibnamefont {Rodriguez}},\
  }\bibfield  {title} {\bibinfo {title} {Scalar potentials for light in a
  cavity},\ }\href {https://doi.org/10.1103/PhysRevResearch.5.013154}
  {\bibfield  {journal} {\bibinfo  {journal} {Phys. Rev. Res.}\ }\textbf
  {\bibinfo {volume} {5}},\ \bibinfo {pages} {013154} (\bibinfo {year}
  {2023})}\BibitemShut {NoStop}%
\bibitem [{\citenamefont {Geng}\ \emph {et~al.}(2020)\citenamefont {Geng},
  \citenamefont {Peters}, \citenamefont {Trichet}, \citenamefont {Malmir},
  \citenamefont {Kolkowski}, \citenamefont {Smith},\ and\ \citenamefont
  {Rodriguez}}]{Geng20}%
  \BibitemOpen
  \bibfield  {author} {\bibinfo {author} {\bibfnamefont {Z.}~\bibnamefont
  {Geng}}, \bibinfo {author} {\bibfnamefont {K.~J.~H.}\ \bibnamefont {Peters}},
  \bibinfo {author} {\bibfnamefont {A.~A.~P.}\ \bibnamefont {Trichet}},
  \bibinfo {author} {\bibfnamefont {K.}~\bibnamefont {Malmir}}, \bibinfo
  {author} {\bibfnamefont {R.}~\bibnamefont {Kolkowski}}, \bibinfo {author}
  {\bibfnamefont {J.~M.}\ \bibnamefont {Smith}},\ and\ \bibinfo {author}
  {\bibfnamefont {S.~R.~K.}\ \bibnamefont {Rodriguez}},\ }\bibfield  {title}
  {\bibinfo {title} {Universal scaling in the dynamic hysteresis, and
  non-{M}arkovian dynamics, of a tunable optical cavity},\ }\href
  {https://doi.org/10.1103/PhysRevLett.124.153603} {\bibfield  {journal}
  {\bibinfo  {journal} {Phys. Rev. Lett.}\ }\textbf {\bibinfo {volume} {124}},\
  \bibinfo {pages} {153603} (\bibinfo {year} {2020})}\BibitemShut {NoStop}%
\bibitem [{\citenamefont {Peters}\ \emph {et~al.}(2021)\citenamefont {Peters},
  \citenamefont {Geng}, \citenamefont {Malmir}, \citenamefont {Smith},\ and\
  \citenamefont {Rodriguez}}]{Peters21}%
  \BibitemOpen
  \bibfield  {author} {\bibinfo {author} {\bibfnamefont {K.~J.~H.}\
  \bibnamefont {Peters}}, \bibinfo {author} {\bibfnamefont {Z.}~\bibnamefont
  {Geng}}, \bibinfo {author} {\bibfnamefont {K.}~\bibnamefont {Malmir}},
  \bibinfo {author} {\bibfnamefont {J.~M.}\ \bibnamefont {Smith}},\ and\
  \bibinfo {author} {\bibfnamefont {S.~R.~K.}\ \bibnamefont {Rodriguez}},\
  }\bibfield  {title} {\bibinfo {title} {Extremely broadband stochastic
  resonance of light and enhanced energy harvesting enabled by memory effects
  in the nonlinear response},\ }\href
  {https://doi.org/10.1103/PhysRevLett.126.213901} {\bibfield  {journal}
  {\bibinfo  {journal} {Phys. Rev. Lett.}\ }\textbf {\bibinfo {volume} {126}},\
  \bibinfo {pages} {213901} (\bibinfo {year} {2021})}\BibitemShut {NoStop}%
\bibitem [{\citenamefont {Kyriienko}\ \emph {et~al.}(2019)\citenamefont
  {Kyriienko}, \citenamefont {Sigurdsson},\ and\ \citenamefont
  {Liew}}]{Liew19}%
  \BibitemOpen
  \bibfield  {author} {\bibinfo {author} {\bibfnamefont {O.}~\bibnamefont
  {Kyriienko}}, \bibinfo {author} {\bibfnamefont {H.}~\bibnamefont
  {Sigurdsson}},\ and\ \bibinfo {author} {\bibfnamefont {T.~C.~H.}\
  \bibnamefont {Liew}},\ }\bibfield  {title} {\bibinfo {title} {Probabilistic
  solving of ${NP}$-hard problems with bistable nonlinear optical networks},\
  }\href {https://doi.org/10.1103/PhysRevB.99.195301} {\bibfield  {journal}
  {\bibinfo  {journal} {Phys. Rev. B}\ }\textbf {\bibinfo {volume} {99}},\
  \bibinfo {pages} {195301} (\bibinfo {year} {2019})}\BibitemShut {NoStop}%
\bibitem [{\citenamefont {Opala}\ \emph {et~al.}(2019)\citenamefont {Opala},
  \citenamefont {Ghosh}, \citenamefont {Liew},\ and\ \citenamefont
  {Matuszewski}}]{Opala19}%
  \BibitemOpen
  \bibfield  {author} {\bibinfo {author} {\bibfnamefont {A.}~\bibnamefont
  {Opala}}, \bibinfo {author} {\bibfnamefont {S.}~\bibnamefont {Ghosh}},
  \bibinfo {author} {\bibfnamefont {T.~C.}\ \bibnamefont {Liew}},\ and\
  \bibinfo {author} {\bibfnamefont {M.}~\bibnamefont {Matuszewski}},\
  }\bibfield  {title} {\bibinfo {title} {Neuromorphic computing in
  {G}inzburg-{L}andau polariton-lattice systems},\ }\href
  {https://doi.org/10.1103/PhysRevApplied.11.064029} {\bibfield  {journal}
  {\bibinfo  {journal} {Phys. Rev. Applied}\ }\textbf {\bibinfo {volume}
  {11}},\ \bibinfo {pages} {064029} (\bibinfo {year} {2019})}\BibitemShut
  {NoStop}%
\bibitem [{\citenamefont {Stroev}\ and\ \citenamefont
  {Berloff}(2023)}]{Stroev23}%
  \BibitemOpen
  \bibfield  {author} {\bibinfo {author} {\bibfnamefont {N.}~\bibnamefont
  {Stroev}}\ and\ \bibinfo {author} {\bibfnamefont {N.~G.}\ \bibnamefont
  {Berloff}},\ }\bibfield  {title} {\bibinfo {title} {Analog photonics
  computing for information processing, inference, and optimization},\ }\href
  {https://doi.org/https://doi.org/10.1002/qute.202300055} {\bibfield
  {journal} {\bibinfo  {journal} {Adv. Quantum Technol.}\ }\textbf {\bibinfo
  {volume} {6}},\ \bibinfo {pages} {2300055} (\bibinfo {year}
  {2023})}\BibitemShut {NoStop}%
\end{thebibliography}

\providecommand{\noopsort}[1]{}\providecommand{\singleletter}[1]{#1}%

\end{document}